\documentclass[pdflatex,sn-mathphys-num]{sn-jnl}% Math and Physical Sciences Numbered Reference Style

\usepackage{graphicx}%
\usepackage{multirow}%
\usepackage{amsmath,amssymb,amsfonts}%
\usepackage{amsthm}%
\usepackage{mathrsfs}%
\usepackage[title]{appendix}%
\usepackage{xcolor}%
\usepackage{textcomp}%
\usepackage{manyfoot}%
\usepackage{booktabs}%
\usepackage{algorithm}%
\usepackage{algorithmicx}%
\usepackage{algpseudocode}%
\usepackage{listings}%
\theoremstyle{thmstyleone}%
\theoremstyle{thmstyletwo}%
\theoremstyle{thmstylethree}%
\begin{document}

\title[Article Title]{SS-DPPN: A self-supervised dual-path foundation model for the generalizable cardiac audio representation}

\author[1]{\fnm{Ummy Maria} \sur{Muna}}\email{maria.muna@bracu.ac.bd}
\equalcont{These authors contributed equally to this work.}

\author*[2]{\fnm{Md Mehedi Hasan} \sur{Shawon}}\email{shawon.cse.bracu@gmail.com}
\equalcont{These authors contributed equally to this work.}

\author[2,3]{\fnm{Md} \sur{Jobayer}}\email{mdjob331@student.liu.se}

\author[2,4]{\fnm{Sumaiya} \sur{Akter}}\email{sakter1@umd.edu}

\author[2,5]{\fnm{Md Rakibul} \sur{Hasan}}\email{rakibul.hasan@curtin.edu.au}

\author[1]{\fnm{Md. Golam Rabiul} \sur{Alam}}\email{rabiul.alam@bracu.ac.bd}

\affil[1]{\orgdiv{Department of Computer Science and Engineering}, \orgname{BRAC University}, \orgaddress{\street{Merul Badda}, \city{Dhaka}, \postcode{1212}, \country{Bangladesh}}}

\affil[2]{\orgdiv{Department of Electricial and Electronic Engineering}, \orgname{BRAC University}, \orgaddress{\street{Merul Badda}, \city{Dhaka}, \postcode{1212}, \country{Bangladesh}}}

\affil[3]{\orgdiv{Department of Biomedical Engineering}, \orgname{Linköping University}, \orgaddress{\city{Linköping}, \postcode{58183}, \country{Sweden}}}

\affil[4]{\orgdiv{Department of Electrical and Computer Engineering}, \orgname{University of Maryland}, \orgaddress{\/Paint Branch Drive{}, \city{College Park}, \postcode{20742}, \state{Maryland}, \country{United States}}}

\affil[5]{\orgdiv{School of Electrical Engineering, Computing and Mathematical Sciences}, \orgname{Curtin University}, \orgaddress{\city{Bentley}, \state{WA 6102}, \country{Australia}}}

\abstract {
    The automated analysis of phonocardiograms is vital for the early diagnosis of cardiovascular disease, yet supervised deep learning is often constrained by the scarcity of expert-annotated data. In this paper, we propose the Self-Supervised Dual-Path Prototypical Network (SS-DPPN), a foundation model for cardiac audio representation and classification from unlabeled data. The framework introduces a dual-path contrastive learning based architecture that simultaneously processes 1D waveforms and 2D spectrograms using a novel hybrid loss. For the downstream task, a metric-learning approach using a Prototypical Network was used that enhances sensitivity and produces well-calibrated and trustworthy predictions. SS-DPPN achieves state-of-the-art performance on four cardiac audio benchmarks. The framework demonstrates exceptional data efficiency with a fully supervised model on three-fold reduction in labeled data. Finally, the learned representations generalize successfully across lung sound classification and heart rate estimation. Our experiments and findings validate SS-DPPN as a robust, reliable, and scalable foundation model for physiological signals.
}


\maketitle

Cardiovascular diseases (CVDs) account for a significant proportion of global mortality and remain the leading cause of death worldwide. To rapidly identify and diagnose the high-risk population, the development of accurate tools is required for early detection \cite{almansouri2024early} \cite{olawade2024advancements}. Phonocardiography (PCG) represents the graphic representation of heart sounds and murmurs \cite{GHAREHBAGHI2019105615}. The acoustic signatures of PCG signals, particularly the presence of murmurs or gallops, provide critical diagnostic clues for a range of cardiac pathologies. A heart murmur is produced by turbulent blood flow through the heart and deviates from the regular heartbeat sound. Cardiac auscultation helps filter out patients with indicative heart disease. However, traditional auscultation is inefficient, time-consuming, and highly dependent on an expert and skilled person. In addition, the accurate identification and interpretation of unusual heart sound patterns can be challenging even for experienced cardiologists. Therefore, automated phonocardiogram (PCG) analysis using Artificial Intelligence (AI) is an important research direction to enable a reliable decision support system~\cite{khan2022cardi,alquran2025deep}.

The application of deep learning-based techniques to phonocardiogram (PCG) signals has shown immense potential for the automated detection of cardiac pathologies. These techniques have demonstrated performance in anomalous heart sound detection that is comparable to that of human experts and showed the effectiveness of deep learning-based techniques in detecting cardiac pathologies from phonocardiogram signals. ~\cite{chorba2021deep, noman2019short, lee2022deep, almanifi2022heartbeat, mcdonald2022detection, walker2022dual, lu2022lightweight}. Supervised methods, including Convolutional Neural Networks (CNNs), Long Short-Term Memory (LSTM) networks, and the attention-based transformer models, have achieved state-of-the-art performance in classification tasks. \cite {al2022lightweight, ren2022deep, cheng2023heart, REN2022100322, LIU202349}. Processing the audio signals in 2D form, such as mel-spectrograms, has further improved the effectiveness of these models. Some studies have indicated that this strategy leads to better accuracy than operating on 1D waveforms alone~\cite{araujo2022maiby, knorr2022using, ding2022classification, lu2022lightweight}. However, the success of supervised approaches is fundamentally constrained by their dependance on large, annotated datasets from experts, which are scarce and expensive to acquire in the medical domain. Furthermore, these supervised learning methods require extensive data preprocessing and feature engineering, with a particular emphasis on adequate time-frequency analysis. 

To overcome this data-supervision bottleneck, Self-Supervised Learning (SSL) has worked as a powerful alternative and effective paradigm, leveraging unlabeled data to learn robust and generalizable representations \cite{ericsson2022self, misra2020self}. Inspired by successes in computer vision and general audio processing (\cite{chen2020simple, al2021clar, gong2022ssast}), SSL has been effectively applied to a range of biosignals, including EEG and ECG. These methods are predominantly based on contrastive learning and masked modelling, where the models are trained to learn invariant representations from augmented views of the same input or by regenerating the masked portion of the input~\cite{oord2018representation, kumar2022contrastive, zhang2021unleashing, baade2022mae, quelennec2025masked}. In the PCG domain specifically, the Listen2YourHeart framework demonstrated the power of this approach, achieving promising performance in murmur detection by applying contrastive learning to the 1D waveform \cite{ballas2022listen2yourheart, ballas2024augmentation}.

Despite these advances, critical gaps remain in the field of biosignal processing. The existing semi-supervised learning (SSL) approach for analysing phonocardiogram signals is the Listen2YourHeart framework, which operates on a single data modality and uses a generic fine-tuning method that does not address clinical class imbalances, resulting in a high false-positive rate~\cite{ballas2022listen2yourheart, ballas2024augmentation}. While this single-domain approach is effective, it fails to capture the complete diagnostic picture available in both the temporal and spectral domains, a limitation that multi-domain and feature-fusion techniques have attempted to address in supervised contexts~\cite{wang2021multi, zhang2024heart}. Furthermore, instance-discriminative contrastive losses like NT-Xent, adopted from SimCLR for Listen2YourHeart \cite{chen2020simple, ballas2022listen2yourheart}, can neglect the global geometric structure of the feature space. Moreover, it lacks the rigorous statistical validation or cross-domain evaluation necessary for its generalisability and reliability.

\begin{figure*}[!ht]
    \centering
    \includegraphics[width=\textwidth]{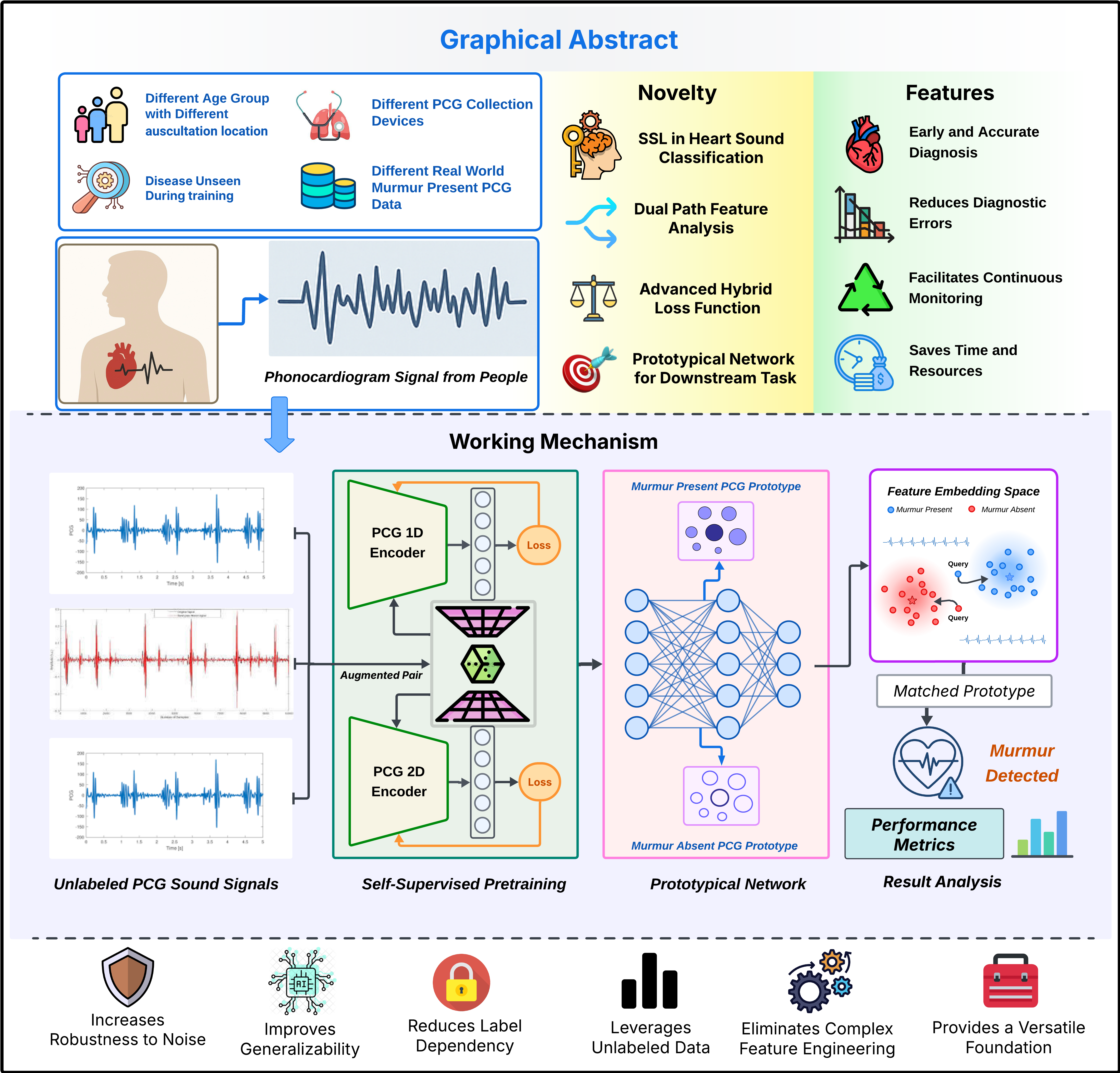}
    \caption{\textbf{Graphical Abstract of SS-DPPN}: Dual-path self-supervised framework for phonocardiogram (PCG). Unlabeled raw PCG signals are augmented and processed by complementary 1D encoder for waveforms and 2D encoder for spectrogram; trained with a hybrid loss; and fine-tuned using a prototypical network. The prototypical network then matches the test embeddings to prototypes for classification. The approach boosts robustness and generalizability across ages, devices, and recording sites, reduces label dependence and feature engineering, and enables early, accurate, and efficient heart-sound screening.\label{fig:graphical_abstract}}  
\end{figure*}

We introduce the Self-Supervised Dual-Path Prototypical Network (SS-DPPN), a novel framework designed to address these limitations and advance the state of the art in three distinct ways. To our knowledge, our dual-path architecture is the first that complementarily learns from both raw and spectrogram heart sounds using enhanced encoders, creating a more holistic feature representation. A dilated temporal convolutional network (TCN) works as an encoder to process the 1D raw waveform to capture temporal dependencies and variations over time~\cite{lea2017temporal}. In parallel, a pre-trained ResNet-50 encoder extracts rich spectro-temporal features from the 2D mel-spectrogram of the same waveform~\cite{he2016deep}. These dual encoders are pre-trained using a contrastive learning method. Second, we introduce a hybrid loss function by combining the instance-discriminative power of NT-Xent loss with the distributional alignment properties of the Wasserstein distance~\cite{panaretos2019statistical}, explicitly encouraging a more robust and well-structured embedding space. The self-supervised model is trained with a range of augmentation strategies to create robust feature representations that are invariant to common acoustic variations. Finally, we incorporate a prototypical network for downstream classification~\cite{snell2017prototypical}, a metric-learning paradigm that yields superior clinical sensitivity with significantly lower false-negative rates. In this work, we demonstrate that SS-DPPN not only sets a new benchmark for heart sound classification but also functions as a versatile and data-efficient foundation model. We also conducted a comprehensive experiment on cross-domain generalization for lung sound classification and heart rate estimation. Finally, rigorous statistical validation confirms the significance and exceptionally well-calibrated performance of our model, establishing its reliability for clinical deployment. Figure~\ref{fig:graphical_abstract} shows the graphical abstract of SS-DPPN.

The main contributions of this paper are as follows:
\begin{itemize}
    \item We introduce a novel self-supervised dual-path framework to create a more comprehensive representation of cardiac audio. By complementarily learning from both 1D waveforms and 2D spectrograms, our approach captures features missed by single-domain methods, which is crucial for robust diagnostics.
    \item We propose an advanced hybrid loss function that creates a more structured and discriminative feature space. By integrating instance discrimination with the distributional alignment of the Wasserstein distance, our loss overcomes the instability of standard contrastive learning to produce more robust and semantically meaningful representations.
    \item We demonstrate that a prototypical network classifier significantly improves diagnostic performance, particularly clinical sensitivity and recall. Standard classifiers often struggle with the class imbalance inherent in medical data, whereas our metric-learning approach significantly reduces false negatives, which is a crucial requirement for an effective clinical screening tool.
    \item We validate the framework as a true foundation model through its exceptional data efficiency and cross-domain generalization across lung sound and heart rate estimation tasks. This directly addresses the critical annotation bottleneck in medical AI and proves the model has learned fundamental bioacoustic features instead of task-specific patterns.
    \item We conduct a rigorous statistical validation, confirming that our model is not only significantly more accurate than baselines but also exceptionally well-calibrated, which is a fundamental aspect for building clinical trust and ensuring model reliability.
\end{itemize}

\section{Results}\label{sec2}
\subsection{Study Overview}

In this section, we present a comprehensive result analysis of our designed approach. We evaluated our self-supervised dual-path prototypical network (SS-DPPN) across four distinct heart sound benchmarks, with all analyses performed at the patient level. A comprehensive statistical, generalizability and reliability analysis was performed on our primary dataset, CirCor 2022\cite{oliveira2021circor}. Model performance was assessed using metrics, including the Area Under the Receiver Operating Characteristic Curve (AUROC), Area Under the Precision-Recall Curve (AUPRC), Accuracy, Precision, Recall, F1-Score, and model calibration scores. Decision thresholds for classification metrics were optimized on a held-out development set and fixed for all subsequent evaluations. Necessary figures and tables are provided for each result section. Other evaluation results are provided in the supplementary document.

\subsection{SS-DPPN Achieves State-of-the-Art Performance across benchmarks}

To establish the credibility and generalizability of our proposed framework, we conducted a rigorous evaluation across four diverse benchmark datasets for heart murmur classification. In this paper, we used the CirCor DigiScope 2022~\cite{oliveira2021circor}, the classic PhysioNet Computing in Cardiology (CinC) Challenge 2016~\cite{Liu_2016}, the Pascal Classifying Heart Sound Challenge (CHSC) 2011~\cite{bentley2011chsc2011}, and the Heart and Lung Sounds Dataset Recorded using Digital Stethoscope (HLS-CMDS) 2025~\cite{torabi2025descriptor} datasets. Figure S1-S4 in the supplementary document, we provided the accuracy and loss curves for training and validation for all the datasets. In addition to that, we all included the corresponding confusion metrices in Figure S5-S8 and ROC and PR curves in Figure S9-S12 in the suplementary document. We considered the largest CirCor DigiScope 2022 dataset the primary one for all other evaluations. As summarized in Table~\ref{tab:combined_comparison_table}(A), our self-supervised method achieves remarkably high and consistent performance across these varied heart sound recordings and clinical settings, demonstrating its ability to handle the complexity and inherent variability in multifaceted recordings.

To establish a broader effectiveness, we performed a comprehensive comparative analysis against some supervised and self-supervised methods on our primary dataset, CirCor digiscope 2022. As presented in Table~\ref{tab:combined_comparison_table}(B), our SS-DPPN framework demonstrates highly competitive performance compared to fully supervised methods. Furthermore, in the scarce field of self-supervised learning for heart sound analysis, our framework also established a foundational benchmark. As shown in Table~\ref{tab:combined_comparison_table}(B), SS-DPPN significantly outperforms the initial work of Ballas et al.\cite{ballas2022listen2yourheart} by a substantial margin, achieving an accuracy of 0.91 and an F1-score of 0.868. Although the overall accuracy was higher compared to their subsequent version, the F1 score was lower because they adopted a combined multi-dataset pre-training strategy with a significantly large number of augmentations, while our model was pre-trained exclusively on a single target dataset with fewer effective augmentation strategies. This highlights the exceptional data efficiency and robustness of our methodology.

A key strength of our approach is the integration of a prototypical network classifier, which demonstrates high sensitivity to the abnormal class (recall = 0.89). In a screening context, this emphasis on sensitivity is clinically critical because it prioritizes the detection of abnormal heart sounds and minimizes false negatives. We therefore accept a modest reduction in precision as an appropriate trade-off, as missing potential pathology is typically more consequential than a false alarm. Notably, we achieve this clinically oriented performance without labeled data during the representation-learning phase, even with highly imbalanced data during validation and testing.

\begin{table}[h!]
\centering
\normalsize
\caption{Overall Model Performance Across Benchmark Datasets and Comparison with Existing Supervised and Self-Supervised Methods on CirCor DigiScope}
\small
\setlength{\tabcolsep}{5pt}
\begin{tabular}{llcccc}
\toprule
 & \textbf{Dataset / Method} & \textbf{Accuracy} & \textbf{Precision} & \textbf{Recall} & \textbf{F1-Score} \\
\midrule
\multirow{4}{*}{\textbf{(A)}} 
 & CirCor DigiScope Dataset~\cite{oliveira2021circor}     & 0.910 & 0.848 & 0.890 & 0.868 \\
 & PhysioNet Challenge Dataset~\cite{Liu_2016}            & 0.881 & 0.948 & 0.898 & 0.922 \\
 & Pascal CHSC Dataset~\cite{bentley2011chsc2011}              & 0.851 & 0.802 & 0.864 & 0.831 \\
 & HLS-CMDS Dataset~\cite{torabi2025descriptor}                & 0.956 & 0.910 & 1.000 & 0.970 \\ 
\midrule
\multirow{8}{*}{\textbf{(B)}} 
 & \multicolumn{5}{l}{\textbf{Supervised Methods}-CirCor Dataset} \\
 & Manshadi et al.~\cite{manshadi2024murmur}           & 0.930 & -     & 0.910 & 0.910 \\
 & Wu et al.~\cite{wu2024heart}                        & 0.736 & 0.730 & 0.830 & 0.768 \\
 & Costandache et al.~\cite{costandache2023automated}  & 0.828 & 0.791 & 0.891 & 0.838 \\
 & Patwa et al.~\cite{patwa2023heart}                  & 0.723 & 0.742 & 0.725 & 0.722 \\
 & \multicolumn{5}{l}{\textbf{Self-Supervised Methods}-CirCor Dataset} \\
 & Ballas et al.~\cite{ballas2022listen2yourheart}     & 0.590 & -     & -     & 0.544 \\
 & Ballas et al.~\cite{ballas2024augmentation}         & 0.836 & -     & -     & 0.900 \\
 & \textbf{SS-DPPN (Ours)}                             & \textbf{0.910} & \textbf{0.848} & \textbf{0.890} & \textbf{0.868} \\
\bottomrule
\end{tabular}
\label{tab:combined_comparison_table}
\end{table}

\subsection{Comprehensive Reliability and Calibration Analysis}

\begin{figure*}[!ht]
    \centering
    \includegraphics[width=\textwidth]{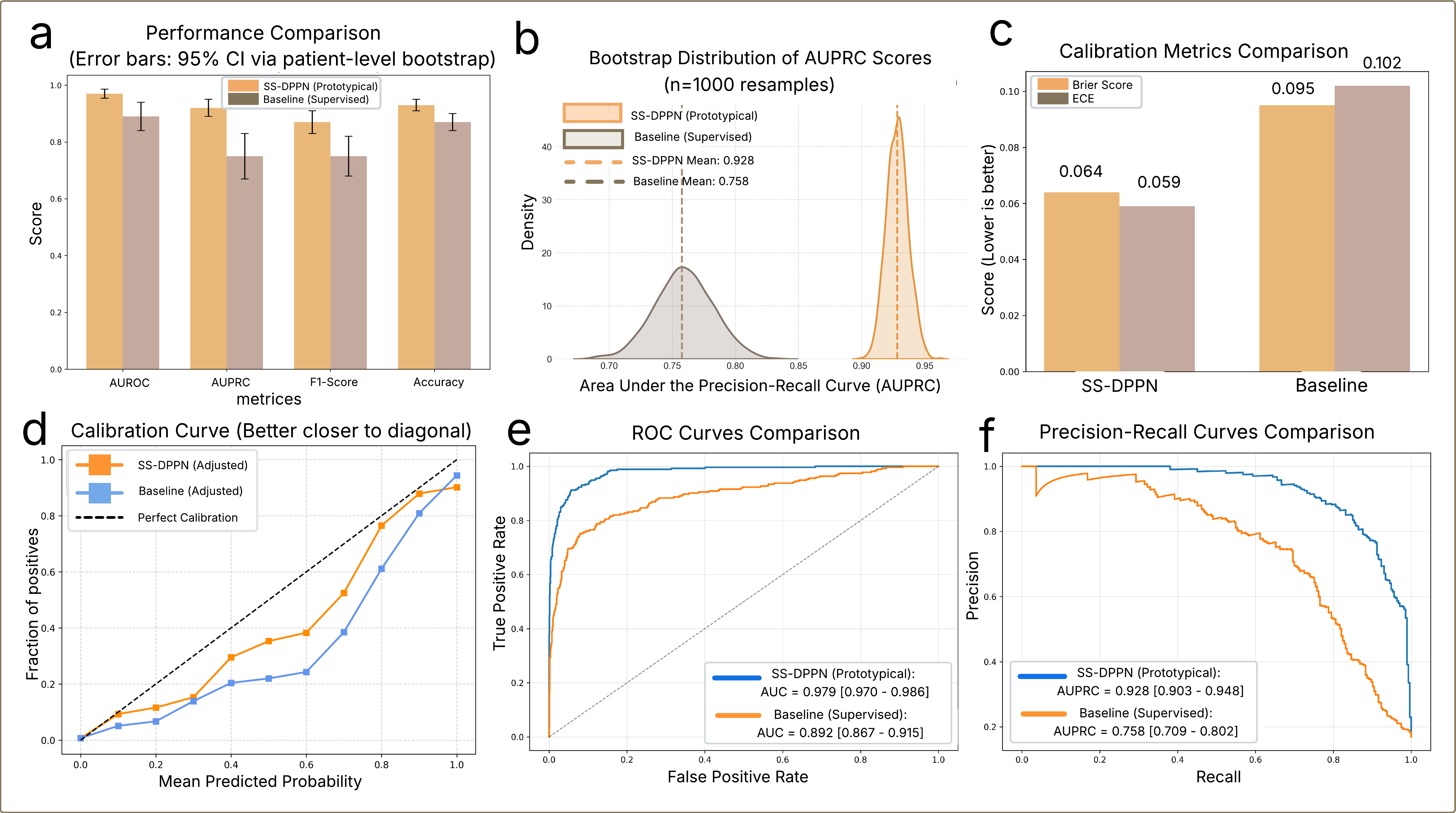}
    \caption{\textbf{Calibration Analysis:} a) Performance Comparison between SS-DPPN and Baseline, b) Bootstrap distribution- changes in AUPRC values across the run, c) Brier and Expected Calibration Error scores of SS-DPPN and Baseline, d) Calibration curve comparison (SS-DPPN closer means better), e) ROC Curve comparison, f) AUPR Curve Comparison. 
    \newline
    \label{fig:calibration}}      
\end{figure*}

To establish the credibility of our framework, we conducted a comprehensive statistical comparison between our SS-DPPN model and a standard supervised baseline model (discussed in the ablation study) in the held-out test set on our primary dataset. SS-DPPN consistently outperformed the baseline on all primary metrics, including accuracy, AUROC, AUPRC, and F1-score, as shown in Figure~\ref{fig:calibration}(a). It shows the non-overlapping 95\% confidence intervals, indicating a high degree of statistical certainty. The robustness is further visualized through the bootstrap distribution of AUPRC scores in Figure~\ref{fig:calibration}(b), where the complete separation and tight distribution of the SS-DPPN model show both its consistent performance(sharper curve) and greater reliability. Since a clinically useful method must provide reliable confidence estimation, we therefore evaluated the calibration of SS-DPPN against the supervised baseline as shown in~\ref{fig:calibration}(d). We found that our framework is more trustworthy, as it is well-calibrated with its confidence scores which closely aligned with its true accuracy, while the baseline is consistently overconfident, with its curve falling farther below the diagonal. The superior Brier Score and Expected Calibration Error (ECE) further quantitatively validate this claim. SS-DPPN significantly generated lower Brier and ECE scores compared to the baseline, as shown in~\ref{fig:calibration}(c), assuring reliability beyond accuracy for clinical interpretation.

Additionally, our SSL model shows statistically superior performance in AUROC, AUPRC, and accuracy through the two-sided DeLong’s test, bootstrap sampling (with replacement) test, and McNemar's test. The significance value (p-value) for the AUROC and accuracy comparisons was both $p < 0.001$, while the significance value for the AUPRC comparison was nearly 0.05 with a two-sided bootstrap test. Although the presence of a class imbalance complicates the PR curves, the observed distributions were clearly separated, proving a consistent improvement. Finally, the performance across all decision thresholds (Figures~\ref{fig:calibration}-e and f) confirms the strong discriminatory power by achieving state-of-the-art AUROC and AUPRC. Overall, these results demonstrate that our SS-DPPN framework is not only significantly more accurate but also more reliable that yields confidence scores that can be meaningfully interpreted in a clinical context.

\subsection{Cross-Domain Transferability}
We also evaluated whether SS-DPPN can generalize to new tasks with different data characteristics beyond its primary learning. These experiments were conducted to prove their foundational capability, which is not specific to heart murmurs but captures patterns from other bioacoustic signals, making them adaptable and versatile.

We first performed a cross-domain transfer learning experiment using the lung sound dataset~\cite{FRAIWAN20211}. The SS-DPPN encoder was first pre-trained exclusively on heart sounds and used as a feature extractor. A prototypical network for the downstream task was minimally fine-tuned on the lung sound data. The model demonstrated exceptional generalization, achieving a diagnostic F1-score of 0.897 and an AUROC of 0.947. Figure~\ref{fig:cross-domain-data-efficiency}(a)(i) displays the tSNE plot of the lung sound data embeddings, which reveal two dense and well-separated clusters representing the 'Normal' and 'Pneumonia' classes, demonstrating exceptional feature separation. This successful transfer, which is achieved with minimal fine-tuning, demonstrates that our model is capable of distinguishing acoustic features applicable across different physiological domains, which is a key characteristic of a foundation model.

To further showcase its versatility, we evaluated the fine-tuned SS-DPPN on an additional regression task, heart rate estimation from ECG signals. The model achieved outstanding performance, with a mean absolute error (MAE) of just 0.9743 BPM and an $R^2$ score of 0.9705, indicating that its predictions account for more than 97\% of the variance in true heart rates. The tight correlation between predicted and true values, as visualised in Figure~\ref{fig:cross-domain-data-efficiency}(a)(ii), confirms that the learned representations are not only useful for classification but also contain precise quantitative information.

\subsection{Pre-training Model Data Efficiency}

\begin{figure*}[!ht]
    \centering
    \includegraphics[width=\textwidth]{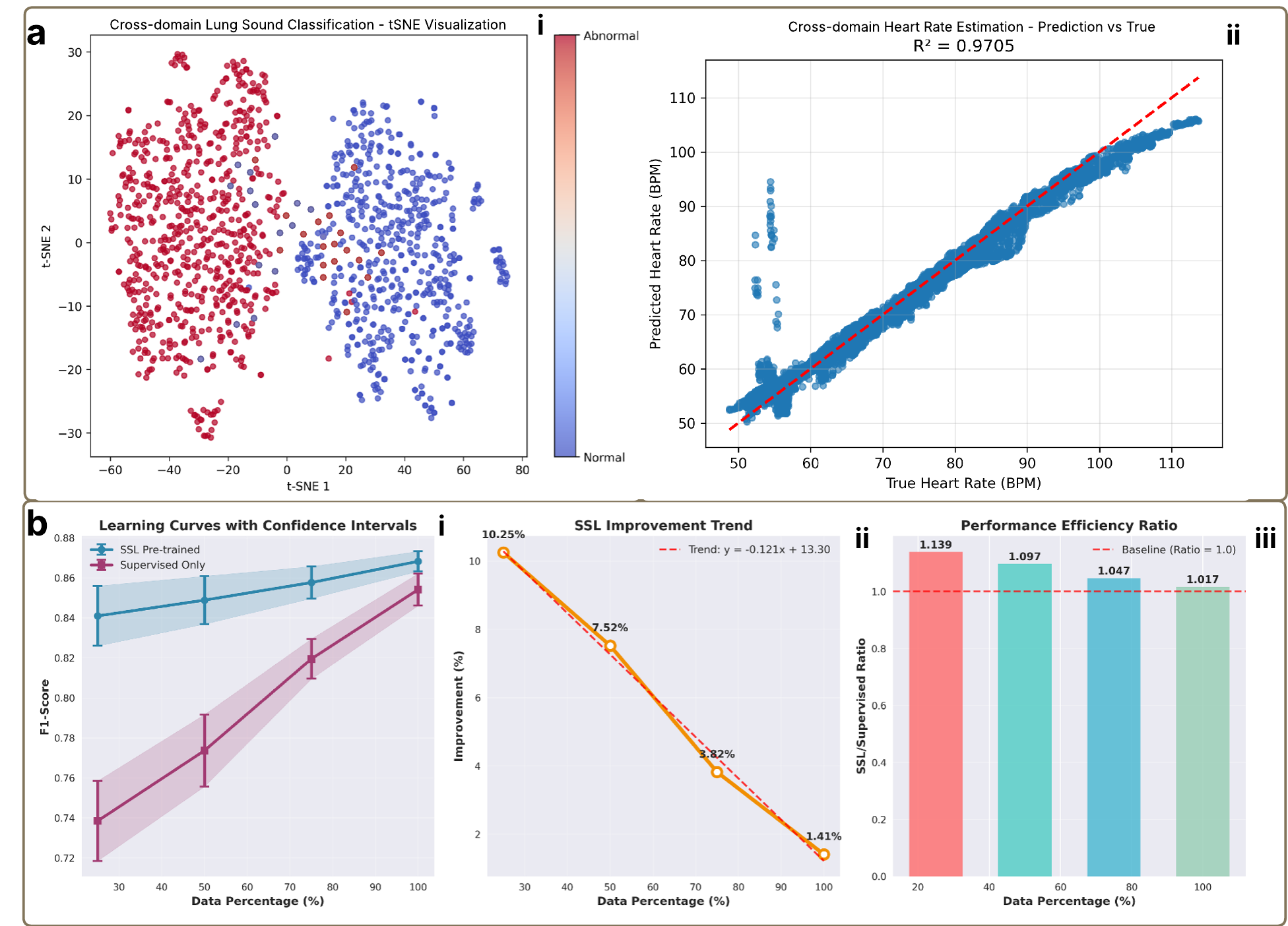}
    \caption{\textbf{a} - tSNE plot for feature separation visualization of lung sound data(i) and prediction vs true class plot for heart rate estimation(ii). \textbf{b} - Comprehensive data efficiency analysis, showing the learning curves with confidence intervals(i), the trend of relative SSL improvement(ii), and the performance efficiency ratio (iii) against a supervised baseline.} 
    \label{fig:cross-domain-data-efficiency}     
\end{figure*}

A central claim of our work is that a foundational model should mitigate the dependency on large, expertly annotated datasets. To quantify the data efficiency gained from our self-supervised pre-training, we compared SS-DPPN against a supervised baseline on progressively smaller subsets of labelled data (Figure~\ref{fig:cross-domain-data-efficiency}-b). The learning curves (Figure~\ref{fig:cross-domain-data-efficiency}- (b-i)) show that our pre-trained model consistently and significantly outperforms the supervised baseline, with the non-overlapping 95\% confidence intervals confirming a statistically robust advantage in low-data regimes. The value of pre-training is most pronounced where labels are scarcest, providing a relative F1-score improvement of over 10\% with just 25\% of the data, a trend that diminishes as more labelled data becomes available (Figure~\ref{fig:cross-domain-data-efficiency}- (b-ii)). Our model achieves a performance level with 25\% of the data that the supervised model only approaches with 75\%, confirming an approximate threefold reduction in the required labelled data and a substantial efficiency gain. The performance efficiency ratio is nearly 1.14 times better in low-data settings (Figure~\ref{fig:cross-domain-data-efficiency}- (b-iii)), validating our framework's potential to address the critical annotation bottleneck in medical AI.

\subsection{Ablation Studies Deconstructs Model Performance}

To rigorously validate our architectural choices and quantify the contribution of each proposed component, we conducted a series of ablation studies. We systematically added or replaced components of the SS-DPPN framework and evaluated its impact on the primary murmur classification task in the CirCor 2022 dataset. First, we show the performance of the single-path baseline model with a standard convolutional network and incrementally introduce the dual-path network, advanced encoder, hybrid loss function, and finally the prototypical network classifier under the umbrella of few-shot learning. The results of this stepwise analysis are summarized in Table \ref{tab:ablation}, confirming that each of our design choices provides a significant and non-trivial contribution to the final performance.

\begin{table}[h!]
\centering
\caption{Ablation Study of Model Components}
\small
\begin{tabular}{p{4.5cm}ccccc}
\toprule
\textbf{Model Config.} & \textbf{Accuracy} & \textbf{Precision} & \textbf{Recall} & \textbf{F1-Score} \\
\midrule
Single Path Base Model \newline (CNN 1D Encoder + Contrastive Learning-NT-Xent Loss + Linear MLP Classifier) 
& 0.805 & 0.749 & 0.704 & 0.725 \\
\midrule
Dual Path Base Model \newline (Single Path Base Model + Mel Spectrogram + 2D CNN Encoder) 
& 0.831 & 0.762 & 0.738 & 0.749 \\
\midrule
Dual Path Enhanced Model \newline (Dual Path Base Model + Dilated 1D Encoder + ResNet 2D Encoder) 
& 0.851 & 0.785 & 0.749 & 0.766 \\
\midrule
Dual Path Enhanced Loss \newline (Dual Path Enhanced Model + Hybrid Loss (NT-Xent Loss + Wasserstein Loss)) 
& 0.863 & 0.792 & 0.758 & 0.774 \\
\midrule
Full Model (SS-DPPN) \newline (Dual Path Enhanced Loss + Prototypical Network) 
& \textbf{0.910} & \textbf{0.848} & \textbf{0.890} & \textbf{0.868} \\
\bottomrule
\end{tabular}
\label{tab:ablation}
\end{table}

The foundation of our model's success lies in the enhanced dual-path encoders and prototypical network in the few-shot paradigm. The basic 1D encoder shows very limited performance and a lower recall score. Introducing a dual-path network enhances its performance by a standard margin, proving the importance of feature integration in the 2D version as well. In the next step, replacing shallow 1D and 2D encoders with a dilated TCN and a pre-trained ResNet-50 provided a further performance lift, validating better long-range temporal and rich spectral feature extraction. Building on this architectural foundation, our hybrid loss function caused a slight performance gain. We consider this minimal contribution an important factor, as the addition of a Wasserstein distance component to the standard NT-Xent loss ensures regularized and global feature space. However, replacing the standard linear classifier with a prototypical network achieved the most pronounced improvement. This change yielded a critical enhancement in the accuracy and particularly the recall value (clinical sensitivity), which rose from 0.758 to 0.890, contributing to the 9\% increase in F1-score. The result confirms the network's superior ability to identify true murmur cases, which is a paramount objective where minimizing false negatives is crucial. Overall, these findings confirm that each component makes a significant, non-trivial contribution and that is the reason behind the model's state-of-the-art performance.

\subsection{Qualitative Analysis of Learned Representations}

\begin{figure}[!ht]
    \centering
    \includegraphics[width=1\linewidth]{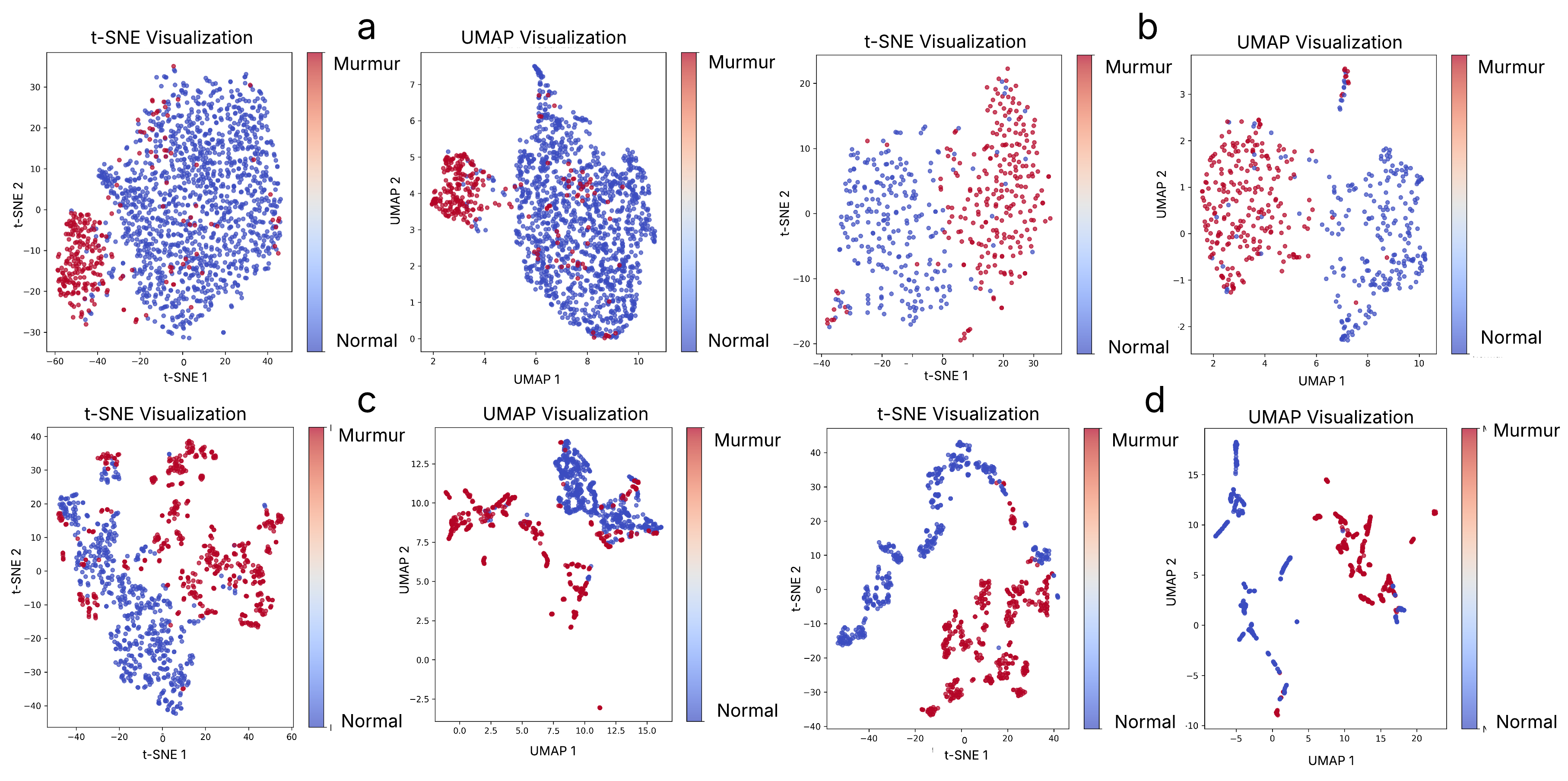}
    \caption{\textbf{t-SNE and UMAP feature separation visualization.} To show the robustness and generalizability of the SS-DPPN, feature separation visualizations for four datasets are added in a, b, c and d, respectively. 
    \newline
    \label{fig:visualization}}      
\end{figure}

We validated the effectiveness of SS-DPPN quantitatively by understanding the learned feature space. We performed feature visualization by projecting high-dimensional embeddings of the test sets from our four benchmark datasets into a 2D space using both t-SNE and UMAP. The results given in Figure~\ref{fig:visualization} demonstrate that our self-supervised pre-training using the dual-path architecture consistently creates a highly structured and semantically meaningful embedding space.

For the CirCor 2022 (Figure~\ref{fig:visualization}-a) and PhysioNet 2016 (Fig.~\ref{fig:visualization}-b) datasets, the embeddings form two dense, distinct, and well-separated clusters corresponding to the 'Murmur' and 'Normal' classes, with minimal overlap. This clear separation provides strong qualitative evidence for the model's high quantitative performance on these large-scale benchmarks. The robustness of the model to class imbalance was particularly proved by the feature separation of the CirCor 2022 test set, where the embeddings for the significantly minority 'murmur' class formed a distinct and well-separated cluster from the significantly larger 'normal' samples. However, the clusters show more overlap and a less defined structure on the PASCAL dataset (Fig.~\ref{fig:visualization}-c), which had also produced the lowest scores in other metrics. This phenomenon is because of its smaller number of real-world clinical recordings and exclusive training settings, which constrains the model from learning a feature space with clear class separation. Despite the smaller number of samples, the model learns a comparatively better and well-separated feature space on the HLS-CMDS dataset than the Pascal dataset (Fig.~\ref{fig:visualization}-d), as the simulated data inherently contains less variability and noise. This visualization supports our quantitative findings and suggests that the poorer performance on this dataset is attributable to a less distinct underlying feature distribution, likely due to differences in data quality or recording protocols.

\section{Discussion}
Our investigation validates that the Self-Supervised Dual-Path Prototypical Network (SS-DPPN) is a powerful and data-efficient framework for cardiac audio analysis that establishes a new benchmark for self-supervised learning in this domain. Our ablation study and the profound statistical analysis confirm that the state-of-the-art performance of SS-DPPN is not incidental but a new trustworthy paradigm for heart murmur detection. The dual-path architecture learns a proper representation of the audio signal, enhanced by our hybrid loss. The prototypical network for the downstream classification task plays one of the most significant roles in the entire architecture. This metric-learning approach demonstrably boosts recall, which has paramount importance for a clinical screening tool, as it corresponds to a lower false-negative rate. The emphasis on sensitivity guarantees the identification of fewer potential pathologies, a reasonable exchange for a minor reduction in precision within a diagnostic setting.

The most significant contribution of this work is the validation of SS-DPPN as a true foundation model for auscultatory signals that is supported by two key findings. First, our framework demonstrates exceptional data efficiency. The SSL model achieves performance with only 25\% of the labeled data that a fully supervised model cannot reach with 75\%. This helps to reduce the dependency on large, expertly annotated datasets, which is a primary bottleneck in medical AI. Second, the model exhibits remarkable generalizability. The successful cross-domain transfer to lung sound classification and a regression-based heart rate estimation task, using an encoder pre-trained exclusively on heart sounds, strongly indicates that SS-DPPN has learned fundamental and transferable bioacoustic features, rather than patterns specific only to cardiac murmurs. 

A key insight we found from our multi-dataset evaluation is the profound influence of data characteristics on model performance. The model achieved state-of-the-art results with the self-supervised approaches. SS-DPPN showed superior performance on the CirCor digiscore 2022 and PhysioNet 2016 real-world datasets and more exceptional performance on the HLS-CMDS dataset, where, despite the small data size, the model excelled at clean, highly consistent signals generated artificially. However, on the older PASCAL dataset, the model's performance was the least compared to others, probably due to the small patient sample and higher signal variability. Another prominent advantage of this SSL technique is that it operates directly on standardized waveforms and does not require the need for complex feature extraction or comprehensive time-frequency analysis. Overall, the framework consistently performs well across diverse conditions, from clean simulated signals to noisy large-scale and imbalanced clinical data, even without extensive feature engineering.

Despite these promising results, we acknowledge several limitations of our model that provide future research direction on SSL-based bioacoustics classification. First, this study was conducted on publicly available, previously collected clinical recordings; therefore, a prospective clinical validation is required to confirm the model's real-world utility. Second, optimization for high sensitivity naturally results in a trade-off with precision, leading to a higher false positive rate that would create false alarms; therefore, it needs to be managed in any practical clinical screening. Third, the dual-path architecture is also computationally intensive due to its complex parallel processing, making it less feasible for real-world deployment. Finally, our transfer learning experiments were limited to the domain of close bioacoustic signals. 

Future work should explore effective single-path model compression and maybe quantization techniques to create more efficient variants suitable for deployment and resource-constrained edge devices. Moreover, the interpretability of separable feature space and the specific predictions should be made. To strengthen the robustness, the model's performance on more distant audio domains should be explored. Future directions could also include multi-modal fusion with extra demographic and electronic health records to create more comprehensive patient models and derive hidden connections with specific diseases. In conclusion, these findings validate our proposed framework and provide a clear roadmap for developing the next generation of practical, robust, and data-efficient diagnostic tools for cardiovascular health.

\section{Methods}
\subsection{Dataset}

The CirCor DigiScope Phonocardiogram Dataset~\cite{oliveira2021circor} v1.0.3 represents a comprehensive collection of heart sound recordings designed for automated murmur detection and classification. The dataset comprises phonocardiogram recordings from 942 patients of age 0-22 years. The dataset includes 5,282 individual heart sound segments with four anatomical auscultation locations, such as the aortic valve (AV), mitral valve (MV), pulmonary valve (PV), and tricuspid valve (TV), from each patient. The recordings are sampled at 4 kHz with variable durations ranging from 4.8 seconds to 80.4 seconds, resulting in more than 33.5 hours of recording. Each segment is annotated with murmur presence labels- "Present," "Absent," or "Unknown"—based on expert cardiologist assessments. The dataset exhibits natural class imbalance, with murmur-present cases representing approximately 30\% of usable recordings. The PhysioNet/Computing in Cardiology Challenge (CinC) 2016 dataset~\cite{Liu_2016}, a collection of 3,126 public phonocardiogram (PCG) recordings. Each file has a duration of 5 to 120 seconds and has been resampled to a frequency of 2,000 Hz. Recordings were from diverse and often noisy environments and were intended for classifying heart sounds as normal or abnormal. The PASCAL Classifying Heart Sounds Challenge 2011 dataset~\cite{bentley2011chsc2011} provides noisy, real-world heart sound audio files that are 1–30 seconds long and come from diverse sources for the task of classifying beats into categories such as normal, murmur, artifact, and extrasystole. Finally, the HLS-CMDS dataset \cite{torabi2025descriptor} provides 535 high-quality, 15-second heart and lung sound recordings from a clinical manikin, uniquely offering individual, mixed, and corresponding source-separated audio files for AI research.

\subsection{Data Preprocessing}
% --------------------------------------------------------
Data preprocessing was designed to produce a patient-aware stratified splitting strategy so that no patient appears in more than one split while maintaining class distribution across training, validation, and test sets. At first, we extracted systolic murmur intervals from the primary dataset using the given start-end timestamps and mapped them to the corresponding PCG recordings. We skipped recordings that do not have a corresponding murmur annotation or have an invalid valve type. The segmentation results for varying lengths of recordings ranged from 0.2 seconds to 3.2 seconds, and the number of murmur-present intervals per patient was substantially lower than the murmur-absent intervals.

To address the variable-length nature of the recordings while preserving natural cardiac timing, we followed a rollover buffer approach that creates a fixed-length 4000-sample window (1 second) by concatenating the adjacent heart sound segments from the same patient. This approach maintains cardiac cycle integrity by employing phase-aware concatenation that maintains natural cardiac cycle progression, validates realistic intervals between the phases, and preserves murmur characteristics across segment boundaries. Moreover, we ensured realistic timing between merged segments by compelling minimum gap thresholds of 50 milliseconds and maximum gap thresholds of 1000 milliseconds, while validating that timing gaps do not exceed 1.5 times the expected cardiac cycle duration based on the estimated heart rate. We further comprehensively assess each rollover window through temporal continuity checks, amplitude consistency analysis, frequency consistency verification, cardiac rhythm preservation assessment, label purity confirmation, and recording source unity validation. 

The patient-aware stratified splitting distributed patients across training (60\%), validation (20\%), and test (20\%) sets while maintaining overall murmur prevalence distribution, preventing data leakage. To address the data imbalance of the train set, we applied a suite of randomized augmentation oversampling for the minority to generate diversity. We employ diverse augmentation strategies, including additive Gaussian noise (0.001 to 0.01), temporal shifting ($\pm 100$ milliseconds), pitch shifting (2 semitones), amplitude scaling ($\pm 20\%$), additive noise with signal-to-noise ratio (SNR) (20-30 dB), and random bandpass filtering with a cutoff frequency (15-25 Hz for low cut and 450-550 Hz for high cut). The augmentation pipeline generates synthetic training samples using combinations of these transformations while maintaining the original validation and test sets in their natural imbalanced state to ensure realistic validation and evaluation conditions. Finally, data normalization is applied per segment using z-score standardization.

\subsection{The SS-DPPN Model Architecture}
\begin{figure*}[!ht]
    \centering
    \includegraphics[width=\textwidth]{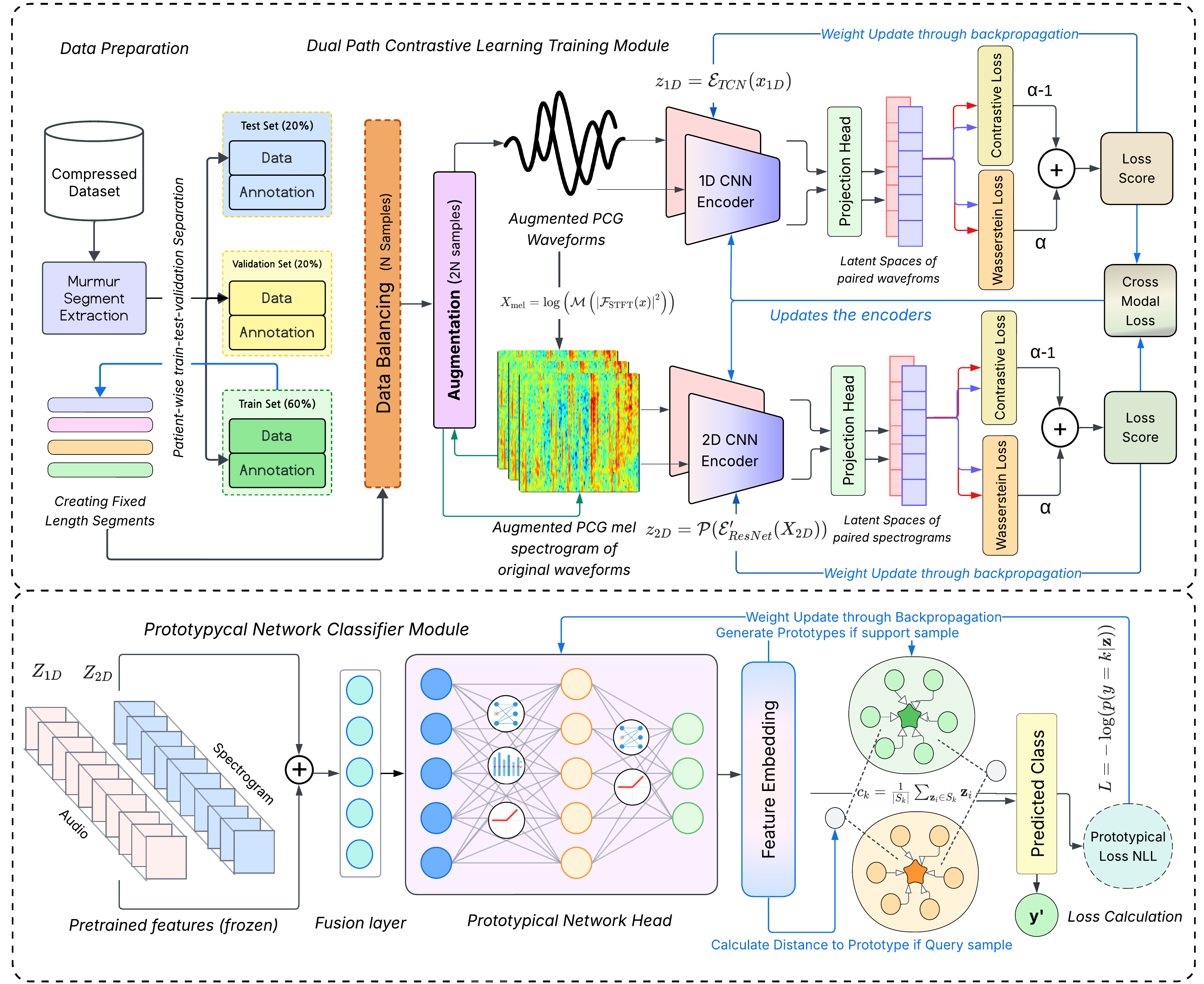}
    \caption{\textbf{Detailed Model Architecture}. The proposed SS-DPPN model for heart sound classification has two major modules: one for self-supervised pretraining using dual path contrastive learning, and another is a prototypical network to calculate the probability of the class label.
    \newline
    \label{fig:overall-model-architecture}}      
\end{figure*}

This paper proposes a comprehensive pipeline designed for automated cardiac murmur detection. The foundation of our methodology is a self-supervised dual-path feature learning strategy, designed to learn robust and discriminative representations of phonocardiogram (PCG) signal without relying on labels. A contrastive learning framework is employed in both paths; the first path processes the raw waveform and preserves high-fidelity temporal information, while the second path processes the mel-spectrogram of the waveform and learns rich spectro-temporal textures of murmurs. The learning is driven by a contrastive framework where augmented views of each sample are created using multiple augmentation strategies. A sophisticated hybrid loss function is employed, combining a distribution-aware Wasserstein loss with a traditional instance-discriminative NT-Xent loss to preserve discrimination and distributional alignment. This loss is uniquely applied not only within each modality but also in a cross-modal fashion (audio-spectrogram). Finally, for the downstream classification task, we employ a prototypical network, a metric-learning approach that is inherently robust to class imbalance. The network computes a single prototype vector for each class and classifies based on their proximity to learned, class-specific prototype centroids in the embedding space. This method is fine-tuned to create a geometrically intuitive and highly separable feature space. Figure~\ref{fig:overall-model-architecture} shows the overall model architecture of SS-DPPN.

\subsubsection{Dual-Path Encoder}
The core of our framework is a dual-path encoder designed to collectively learn from both the 1D temporal waveform and its 2D spectrogram representation. The two paths process these inputs in parallel using a dilated Temporal Convolutional Network (TCN) to process the 1D input signal and a pre-trained ResNet to process the 2D mel-spectrogram. This architecture allows the model to fuse complementary information from both domains, creating a more rigorous representation of the cardiac audio signal.

% \subsection{Dilated Temporal Convolutional Network Encoder for Raw Waveform}
\textbf{Dilated Temporal Convolutional Network 1D signal Encoder.} We designed an encoder based on a Dilated Temporal Convolutional Network (TCN) to process the raw 1D audio waveform and capture a wide range of temporal dependencies. This architecture is specifically suitable for time-series data as it uses dilated convolutions to exponentially increase the receptive field without an increase in parameters, enabling the efficient modeling of long-term patterns in the phonocardiogram. 

The encoder consists of an initial downsampling block, a stack of eight residual dilated blocks, and a final pooling layer to produce a fixed-dimension embedding. The hierarchical forward pass of the 1D encoder is shown in Equation~\ref{eq:tcn1}. An input waveform $X$, is first processed by an initial convolutional block \(f_{initial}\) comprising a lightweight 1D convolution with a large kernel and stride, followed by Layer Normalization, a ReLU activation, and average pooling. The resulting feature map \(L_0\), is then passed through a composition of eight sequential residual dilated blocks \(f_{block,l}\). Finally, a terminal adaptive average pooling layer \(f_{pool}\) aggregates the features from the final block \(h_8\), into a single fixed-dimension embedding, \(Z_{1D}\).
\begin{equation}
\begin{aligned}
\mathbf{h}_0 &= f_{\text{initial}}(\mathbf{x}) \\
\mathbf{h}_{l+1} &= f_{\text{block}, l}(\mathbf{h}_l) \quad \text{for } l=0, \dots, 7 \\
\mathbf{z}_{\text{1D}} &= f_{\text{pool}}(\mathbf{h}_8)
\end{aligned}
\label{eq:tcn1}
\end{equation}

Each residual dilated block is designed for stable training in a deep network. It employs pre-activation normalization layer, which receives the input \(x_l\) from the block and passed to a ReLU activation function. The result is then processed by a single 1D convolution with an exponentially increasing dilation factor (\(d_l=2^l\)) and a subsequent dropout layer. The important feature of the block is the identity skip connection, where the original input \(x_l\) is added to the output of the convolutional path. This residual mechanism ensures robust gradient flow and promotes the learning of complex temporal features. The operation within a single block $l$ to produce the output \(x_{l+1}\) is described in Equation~\ref{eq:tcn2}.

\begin{equation}
\begin{aligned}
\mathbf{x}_{\text{pre}} &= \mathrm{ReLU}(\mathrm{BatchNorm1D}(\mathbf{x}_l)) \\
\mathbf{x}_{\text{conv}} &= \mathrm{Dropout}\!\left(\mathrm{Conv1D}_{\text{dilated}}(\mathbf{x}_{\text{pre}})\right) \\
\mathbf{x}_{l+1} &= \mathbf{x}_{\text{conv}} + \mathbf{x}_l
\label{eq:tcn2}
\end{aligned}
\end{equation}

\textbf{ResNet-50 2D Spectrogram Encoder.} We leverage a pre-trained ResNet-50 architecture to process the 2D spectro-temporal features and extract powerful hierarchical features from log-mel spectrograms derived from the audio signals. As the spectrograms are single-channel, the ResNet-50 model's primary convolutional layer was modified to accept a single input channel instead of three, and the weights were initialized by averaging the weights of the original three-channel pre-trained layer. The ResNet backbone is used purely as a feature extractor, with its final fully connected classification layer removed. The fundamental component of the ResNet is the residual block. For a 2D feature map \(X_l\), the output \(X_{l+1}\) is generated via a skip connection, as shown in Equation~\ref{eq:resnet-feature-map-final}:

\begin{equation}
\mathbf{X}_{l+1} = \text{ReLU}(\mathcal{G}(\mathbf{X}_l, {W_i}) + \mathbf{X}_l)
\label{eq:resnet-feature-map-final}
\end{equation}

where $\mathcal{G}$ represents the residual function, which is a stack of 2D convolutions within the block, and \({W_i}\) are the learnable weights. The high-dimensional feature vector from the ResNet backbone is then passed through a projection head, which consists of a two-layer Multi-Layer Perceptron (MLP) with BatchNorm and Dropout. This head maps the features to the final lower-dimensional embedding, tailoring them for the specific task of murmur detection. The operation of the projection head, $P$, on an input feature vector $h$ is defined as:

\begin{equation}
\mathcal{P}(\mathbf{h}) = W_p \cdot \mathrm{ReLU}\left( \mathrm{BN}(W_e \mathbf{h}) \right)
\label{eq:projection-head-final}
\end{equation}

where \(W_e\) and \(W_p\) are the weight matrices of the two linear layers. The complete forward pass of the encoder to produce the final 2D feature embedding \(Z_{2D}\) from an input spectrogram $X$ is given in Equation~\ref{eq:2D-feature-embedding-final}:

\begin{equation}
\mathbf{z}{2D} = \mathcal{P}\left( \mathcal{F}'{\text{ResNet-50}}(\mathbf{X}) \right)
\label{eq:2D-feature-embedding-final}
\end{equation}

where $F'_{ResNet-50}(·)$ is the modified ResNet-50 feature extractor function.

% \subsection{Feature Fusion}
\textbf{Feature Fusion.} The feature fusion mechanism is designed to jointly integrate the complementary representations learned by the parallel 1D temporal and 2D spectro-temporal encoders. The feature vector from the 1D encoder, \( \mathbf{z}_{1D} \in \mathbb{R}^D \), and the vector from the 2D encoder, \( \mathbf{z}_{2D} \in \mathbb{R}^D \), are first concatenated to form a single vector. This combined vector is then passed through a fusion layer, \( \mathcal{F}_{\text{fusion}} \), a multi-layer perceptron (MLP), which learn the optimal non-linear combination of these features. The resulting unified feature vector is \( \mathbf{Z}_{\text{fused}} \) which is given in equation \ref{eq:fusion}:
\begin{equation}
    \mathbf{Z}_{\text{fused}} = \mathcal{F}_{\text{fusion}}\left( [\, \mathbf{z}_{1D} , \mathbf{z}_{2D} \,] \right)
    \label{eq:fusion}
\end{equation}

The advantage of this feature fusion approach is the creation of a comprehensive representation that is more informative than either modality alone. It allows the model to understand precise temporal information with the complex spectral textures of murmurs and enhances the model's discriminative power.

\textbf{Data Augmentation for Contrastive Representation Learning}. A critical component of our self-supervised contrastive learning pipeline is the data augmentation strategy, which generates semantically consistent but distinct views of each input sample. The objective is to teach the model invariance to common acoustic and temporal variations while forcing it to learn the core, underlying features that define a heart murmur sound. To align with our dual-path architecture, we apply a set of modality-specific augmentations to both the 1D waveform and its 2D spectrogram representation. We further applied augmentation on the spectrogram because it introduces a completely different and complementary set of challenges for the 2D encoder, forcing it to become more robust in ways that 1D waveform augmentation alone cannot achieve.

\textbf{(a) 1D Waveform Augmentations}: To create a powerful contrastive learning encoder, we apply a series of augmentations directly to the raw audio waveform. For each sample, a random composition of the following transformations is used:
\begin{itemize}
    \item \textbf{Gaussian Noise Addition:} Additive white Gaussian noise is introduced to the signal to recognize recordings with electronic sensor noise and improve the  robustness to low signal-to-noise ratio environments.
    \item \textbf{Random Time Shifting:} The entire waveform is spherically shifted by a random amount. These modified signals with shuffling, force the model to recognize murmur regardless of their precise temporal alignment within the testing boundary.
    \item \textbf{Amplitude Scaling:} The global amplitude of the waveform is multiplied by a random scalar. This process ensures that the learned features are invariant to variations in recording volume.
    \item \textbf{Time Stretching and Pitch Shifting:}  To moderately stretch the signal duration and shift its pitch, we apply signal processing techniques in the time domain based on resampling algorithms. These transformations encourage the model to learn fundamental acoustic patterns rather than memorizing superficial temporal and frequency characteristics.
\end{itemize}

\textbf{(b) 2-D Spectrogram Augmentations}: For the 2D contrastive pretraining encoder, we employ augmentations for operating on the log-mel spectrogram representation. The augmentations are two distinct masking operations:
\begin{itemize}
    \item \textbf{Time Masking:} A random number of consecutive vertical time steps in the spectrogram are masked or removed. The model is forced to infer features from the surrounding temporal context of the signals.
    \item \textbf{Frequency Masking:} Similar to the time masking, a random number of consecutive horizontal mel-frequency bins are masked. This again occludes specific frequency bands and prevents the model from becoming overly reliant on any single frequency component.
\end{itemize}
The random application of this comprehensive set of augmentations ensures that the two views generated for each heart sound sample in the contrastive learning framework are sufficiently diverse and become a challenging, enabling effective self-supervised learning task.

\subsubsection{Hybrid Contrastive Loss Function}
To train the self-supervised pre-training model using contrastive learning framework, we employ a hybrid loss function for the encoders that combines an instance discriminative contrastive loss and a distribution-matching Wasserstein loss. This hybrid loss function is designed to create a feature representation that is not only highly discriminative at the instance level but also well-structured at the global level, leading to more robust and generalizable representations.

% \subsubsection{Normalized Temperature-scaled Cross-Entropy (NT-Xent)}
\textbf{Normalized Temperature-scaled Cross-Entropy (NT-Xent).} The first component of our hybrid loss is the Normalized Temperature-scaled Cross-Entropy (NT-Xent) loss, which is an essential building block and central to our contrastive learning framework. For a positive pair (sample with its augmented version) of projected embeddings \( (\mathbf{z}_i, \mathbf{z}_j) \) within a batch of \( 2N \) total embeddings, containing \( N \) number of total samples and \( N \) a number of augmented samples, the NT-Xent loss encourages the model to maximize the similarity of the positive pair relative to all other \( 2N - 1 \) negative pairs. The loss for each pair is calculated using:
\begin{equation}
\mathcal{L}_{\text{NT-Xent}}(\mathbf{z}_i, \mathbf{z}_j) = -\log \frac{\exp\left( \text{sim}(\mathbf{z}_i, \mathbf{z}_j) / \tau \right)}{\sum_{k=1}^{2N} {1}_{[k \neq i]} \exp\left( \text{sim}(\mathbf{z}_i, \mathbf{z}_k) / \tau \right)}
\label{eq:NT-Xent-Loss}
\end{equation}
where \( \text{sim}(z_i, z_j) \) denotes the cosine similarity between the sample pairs and \( \tau \) is a temperature hyperparameter that controls the separation of negative samples. The numerator maximizes the cosine similarity \( \text{sim}(z_i, z_j) \) between a positive pair, and the denominator normalizes this value against the sum of similarities between the embedding \( z_i \) and all other \( 2N - 1 \) embeddings in the batch. This effectively pushes \( z_i \) away from negative samples and pulls it closer to its positive counterpart.

% \subsubsection{Wasserstein Metric Loss}
\textbf{Wasserstein Metric Loss.} Although powerful in learning discriminative features, the NT-Xent loss focuses primarily on local relationships between individual samples. Therefore, to ensure that the global structure of the embedding distributions for positive pairs is well-aligned, we incorporate the squared 2-Wasserstein distance as an additional loss term to enhance representation learning. The squared 2-Wasserstein distance, denoted \( W_2^2 \), measures the optimal transport cost to move the probability mass from one embedding distribution \( \mu \) to another \( \nu \). For distributions of embeddings \( \mathbf{z}_a\) denoted as \(\mu \) and for \( \mathbf{z}_b\) denoted as \(\nu\) in \( \mathbb{R}^D \), the Wasserstein loss \(\mathcal{L}_W(\mu, \nu)\) is calculated as:
\begin{equation}
W_2^2(\mu, \nu)=\mathcal{L}_W(\mu, \nu) = \inf_{\gamma \in \Gamma(\mu, \nu)} \int \| \mathbf{z}_a - \mathbf{z}_b \|^2 \, d\gamma(\mathbf{z}_a, \mathbf{z}_b)
\label{eq:wasserstein-loss}
\end{equation}
where the integral \(\int_{\mathbb{R}^D \times \mathbb{R}^D}\) calculates the expected cost for any given plan, using the squared Euclidean distance, \( \lvert \lvert \mathbf{z}_a - \mathbf{z}_b \lvert \rvert^2 \) to move an embedding vector \(\mathbf{z}_a \in \mu\) to \( \mathbf{z}_b \in \nu \) in D-dimensional euclidean space \( \mathbb{R}^D \). \( \inf \) represents the infimum that finds the greatest lower bound or the most efficient plan \( \gamma \) from the entire set of plans \( \Gamma(\mu, \nu) \) that minimizes the total transport cost. This computationally intensive distance is efficiently approximated using the Sinkhorn algorithm provided by the \texttt{GeomLoss} library. This loss component encourages holistic alignment of the feature distributions, offering a smoother and more stable learning signal than instance-wise contrastive loss alone.

% \subsubsection{Hybrid Loss Function}
\textbf{Hybrid Loss Function.} Although the NT-Xent loss function used in the contrastive learning framework exhibits strong discriminative power, it is further increased with the Wasserstein loss to enable more stable training in the CL model and create a better representation learning. The hybrid loss \( \mathcal{L}_{\text{Hybrid}} \) is a weighted linear combination of the two components, balanced by a hyperparameter \( \alpha \):
\begin{equation}
\mathcal{L}_{\text{Hybrid}} = \alpha \mathcal{L}_W + (1 - \alpha) \mathcal{L}_{\text{NT-Xent}}
\label{eq:hybrid-loss}
\end{equation}
where \( \alpha \) is a scalar hyperparameter, which acts as a weighting factor to balance the contributions of the two loss functions. In our experiments, we tuned and set \( \alpha = 0.3 \) to make the learning process focus primarily on the powerful contrastive objective while moderately using the Wasserstein distance as a geometric regularizer to encourage a more globally structured and stable embedding space. 
%This hybrid approach results in a more structured and effective final representation for downstream tasks.

\subsubsection{Prototypical Network Classifier}
We propose a different technique for the downstream task instead of a linear classifier. Since our validation and test sets of our primary dataset are heavily skewed with significantly fewer murmur-present segments, we employ a Prototypical Network that operates in a metric space and performs classification based on distances to class prototypes learned as the mean embeddings of support samples, making them naturally more robust to imbalance and better suited for few-shot scenarios. Traditional classifiers, such as softmax-based linear heads, often struggle in such settings due to their reliance on decision boundaries learned from oversampled or synthetic data, while the metric-learning paradigm is different from traditional discriminative classification. Instead of learning a complex, high-dimensional decision boundary to separate classes, a prototypical network learns to model the probability distribution of each class directly within a carefully constructed embedding space. After the self-supervised pre-training, a semantically rich and geometrically structured feature space is already produced, where samples from the same class are naturally co-located. Therefore, the objective is not to fundamentally alter this space; however, to learn an optimal metric within it by identifying a single prototype or centroid for each class and classifying new samples based on their proximity to these prototypes. This is a few-shot learning approach, which is advantageous for its data efficiency and potential for improved robustness and interpretability.

The prototypical network feeds features generated by the frozen dual-path feature extractor. The unified feature vector $\mathbf{z}_{\text{fused}}$ is passed through a dedicated and trainable prototypical head. This head is an MLP that performs a nonlinear transformation and maps the features into a final, lower-dimensional space. 
%The projection head further refines the feature space and learn a transformation that maximizes inter-class distance while minimizing intra-class distance.

The network is further trained using an episodic methodology where each mini-batch is structured as a distinct learning episode \cite{yu2020episode}. Each training batch is partitioned into two disjoint sets, including a support set ($S$) and a query set ($Q$). The support set provides the prototype from which class representations are built, while the query set provides the samples used for loss calculation and optimization by measuring the proximity distance with the prototypes. Subsequently, a single prototype vector $\mathbf{p}_c$ is computed for each class $c$ present in the support set. This prototype represents the centroid of the class within the embedding space and is calculated as the arithmetic mean of the embeddings of all support samples belonging to that class as:
\begin{equation}
\mathbf{p}_c = \frac{1}{|S_c|} \sum_{(\mathbf{x}_i, y_i) \in S_c} f_\phi(\mathbf{x}_i)
\label{eq:prototype}
\end{equation}
where \( S_c \) is the set of support samples belonging to class \( c \), and each sample \( (\mathbf{x}_i, y_i) \in S_c \) consists of an input \( \mathbf{x}_i \) and its corresponding class label \( y_i \), such that \( y_i = c \). \( f_\phi \) is the embedding function parameterized by \( \phi \), and \( f_\phi(\mathbf{x}_i) \) denotes the latent representation of the input \( \mathbf{x}_i \), which is used to calculate the prototype \( \mathbf{p}_c \) for class \( c \).

A query point \(X_q\) is passed through the embedding function \(f_\phi\), and then a probability distribution is computed across all class prototypes \(\mathbf{p}_c\). This distribution is a softmax function over the negative squared Euclidean distances between embedding of the query sample and each prototype. The closest prototype determines the predicted class. The probability that \(X_q\) belongs to class c is given by:
\begin{equation}
p_\phi(y=c \mid \mathbf{x}_q) = \frac{\exp(-\|f_\phi(\mathbf{x}_q) - \mathbf{p}_c\|^2)}{\sum_{c' \in C} \exp(-\|f_\phi(\mathbf{x}_q) - \mathbf{p}_{c'}\|^2)}
\label{eq:query}
\end{equation}
where C is the set of all classes in the support set, \(f_\phi(\mathbf{x}_q)\) represents the embedding of the query sample, \(\mathbf{p}_c\) is the mean embedding of class c, \(-\|f_\phi(\mathbf{x}_q) - \mathbf{p}_{c'}\|^2\) is the squared euclidean distance between the query and class prototype, and the denominator is the normalization across all classes.

The parameters of the projection head \( \phi \) are optimized by minimizing the negative log-likelihood over the query set. The loss function used to train the Prototypical Network is defined as:
\begin{equation}
\mathcal{L}_{\text{Proto}} = \sum_{(\mathbf{x}_q, y_q) \in Q} -\log p_\phi(y_q \mid \mathbf{x}_q)
\label{eq:proto-loss}
\end{equation}
where \( Q \) is the query set, and \( p_\phi(y_q \mid \mathbf{x}_q) \) is the probability that the query sample \( \mathbf{x}_q \) is correctly classified into class \( y_q \) based on its distance to class prototypes in the embedding space.

At inference time, prototypes for each class are first computed using a representative support set of labeled examples from the training data. An unseen test sample is then passed through the complete embedding function \(f_\phi(.)\). Its class is determined by finding the prototype to which its embedding has the minimum Euclidean distance, constituting a simple and efficient nearest-centroid decision rule.

\subsection{Experimental Setup and Training Procedures}
% \subsubsection{Self-Supervised Pre-training}
The SS-DPPN model and all its experiments were implemented using the PyTorch framework and trained on an NVIDIA GeForce RTX 3060 Ti GPU. All random seeds for PyTorch, NumPy, and CUDA were fixed to 42 for complete reproducibility. The initial dual-path self-supervised pre-training was conducted for 30 epochs with a batch size of 32, using the Adam optimizer and a Cosine Annealing learning rate schedule starting at \( 1 \times 10^{-3} \). The hybrid contrastive loss was configured with a temperature \( \tau = 0.07 \) and a Wasserstein distance weighting factor of \(\alpha = 0.3\).

% \subsubsection{Downstream Task Fine-tuning}
 Following pre-training, our proposed Prototypical Network was trained for 50 epochs by learning the optimal mapping to the prototypical space. For this stage, the feature extractor backbone was kept frozen and the projection head was optimized with Adam using a learning rate of \( 1 \times 10^{-4} \), a weight decay of \( 1 \times 10^{-4} \), and gradient clipping with a maximum norm of 1.0 to ensure stability. For comparison, a baseline classifier was also fine-tuned on top of the pre-trained backbone for 50 epochs using an Adam optimizer with a differential learning rate strategy \( 1 \times 10^{-4} \) for the new classifier layers and \( 1 \times 10^{-5} \) for the backbone layers, which was dynamically adjusted by a ReduceLROnPlateau scheduler. During this baseline fine-tuning, the backbone was kept frozen for the initial 10 epochs.

\subsection{Evaluation Metrics and Statistical Analysis}
We employed comprehensive evaluation metrics and rigorous statistical methodologies specifically designed for clinical heart sound classification tasks. Specifically, F1-score, area under the receiver operating characteristic curve (AUROC), and area under the precision recall curve (AUPRC) serve as the significant discrimination metrics. AUROC measures the threshold-independent model performance across all classification boundaries, while the AUPRC provides a complementary assessment particularly valuable in class-imbalanced medical scenarios where minimizing false positives is critical for clinical acceptance. Expected Calibration Error (ECE) and Brier Score were generated for model calibration assessment and evaluation alignment between predicted probabilities and actual outcome frequencies. Reliability visualization for monitoring biases across probability ranges. We also applied the DeLong method for AUROC comparisons, statistical significance testing, McNemar's test for paired classification accuracy differences, and bootstrap hypothesis testing for AUPRC comparisons through patient-level resamples to construct empirical null distributions. Finally, we employ the t-SNE and UMAP dimensionality reduction techniques that visualize high-dimensional features by projecting them into low-dimensional space while preserving structure and similarity. All these metrics ensure robust evaluation of model reliability and clinical applicability across diverse healthcare settings.

\clearpage

\renewcommand{\thefigure}{S\arabic{figure}}
\setcounter{figure}{0}

% \begin{document}

\begin{center}
    \LARGE\textbf{Supplementary Information}
\end{center}

\vspace{1em}

% \section{Graphical Abstract}

% To understand the overall methodology of the article, we have created a graphical abstract:
% \begin{figure}[h!]
%     \centering
%     \includegraphics[width=\textwidth]{GA.png}
%     \caption{Graphical Abstract of the proposed method}
%     \label{fig:graphical_abstract}
% \end{figure}

\section{Result Analysis}

\subsection{Train Validation Loss and Accuracy Curves}
We have added all the loss and accuracy curves. All the curves were smoothed using a 1D Gaussian filter. The number of epochs was set to the point at which the validation curve tends to flatten; that’s why it varies across datasets. Although the training and validation sets were exclusive, the validation almost aligns with the accuracy curves, proving its learning capabilities across the epochs.

\begin{figure}[ht!]
    \centering
    \includegraphics[width=\textwidth]{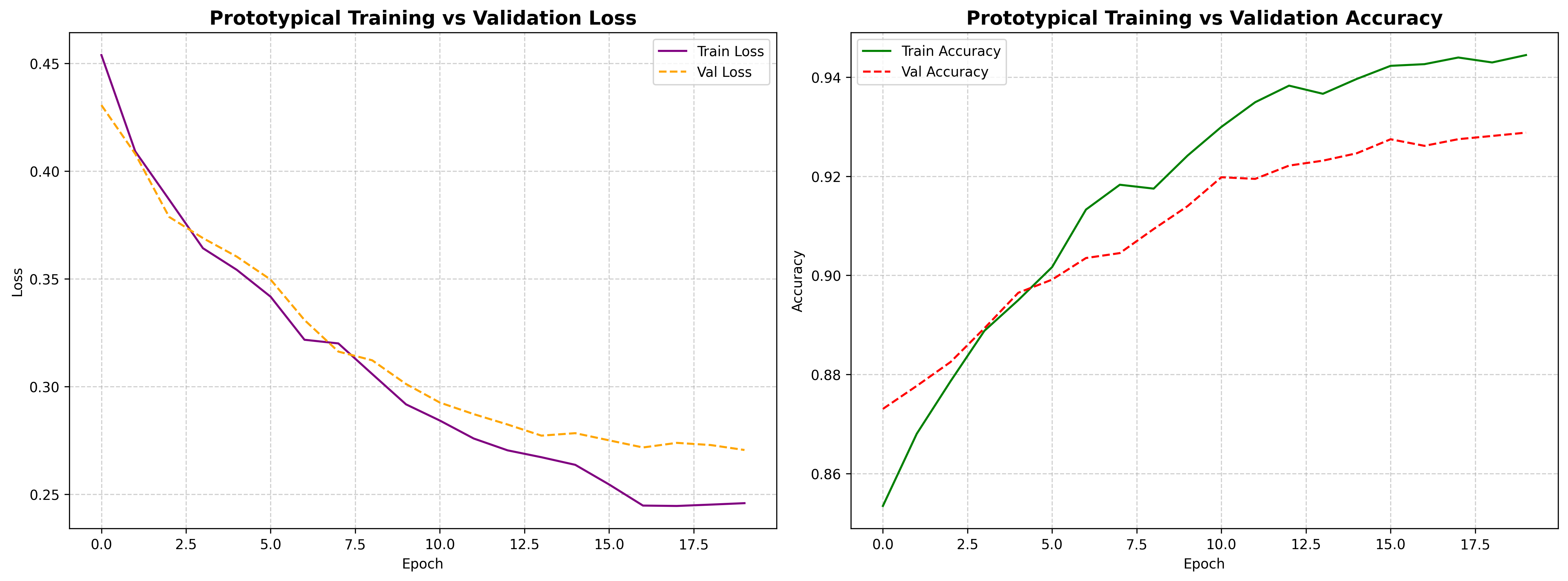}
    \caption{Train-validation loss and accuracy curves for the CirCor 2022 dataset}
    \label{fig:loss_accuracy_curves}
\end{figure}

For the \textbf{CirCor 2022 dataset}, the model achieved validation accuracy of 93\% and 95\% for the training set on 20 epochs. On the test set, it achieved 91\% (as reported in the manuscript-Table 1). This dataset was highly imbalanced, and the validation and test sets contained samples with a natural distribution and quantity proportional to the original positive and negative samples. Therefore, the validation and test sets were also highly imbalanced. Our prototypical network played a great role in identifying the significantly lower murmur present in the class.

\begin{figure}[ht!]
    \centering
    \includegraphics[width=\textwidth]{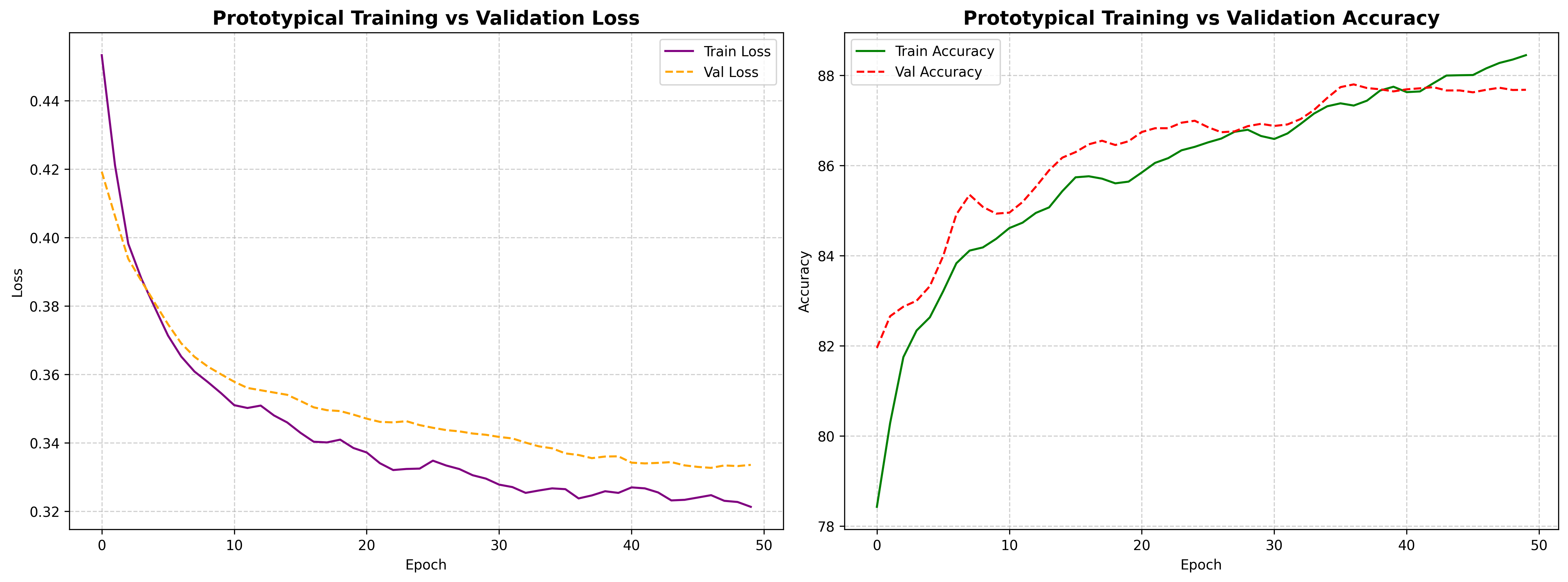}
    \caption{Train-validation loss and accuracy curves for the CinC 2016 dataset}
    \label{fig:loss_accuracy_curves}
\end{figure}

For the \textbf{CinC 2016 dataset}, the model achieved an accuracy greater than 87\% for the validation set and greater than 88\% for the training set for 50 epochs. Higher epochs do not contribute to greater scores than this. The validation score was almost consistent with the training scores. On the test set, it achieved 88.1\% (as reported in the manuscript-Table 1). 

\begin{figure}[ht!]
    \centering
    \includegraphics[width=\textwidth]{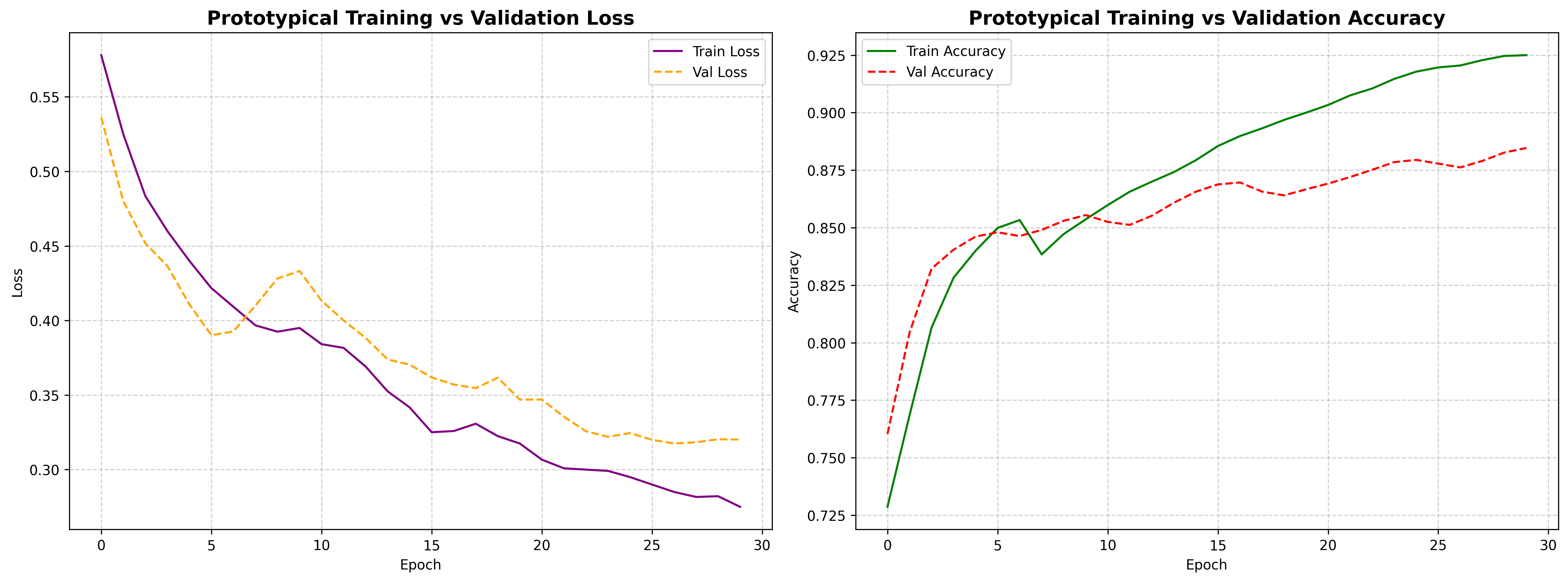}
    \caption{Train-validation loss and accuracy curves for the Pascal dataset}
    \label{fig:loss_accuracy_curves}
\end{figure}

\newpage

For the \textbf{Pascal dataset}, the model achieved accuracy greater than 88\% for the validation set and greater than 92\% for the training set. On the test set, it achieved 85\% (as reported in the manuscript-Table 1). This dataset has shown it tends to overfit. Although the model yielded better results in training, it performed lower in both the validation set and the test set. After extensive hyperparameter tuning and numerous trial-and-error attempts, this was the highest score we could achieve for both the validation and test sets. For this dataset, the optimal number of epochs was 30.

\begin{figure}[ht!]
    \centering
    \includegraphics[width=\textwidth]{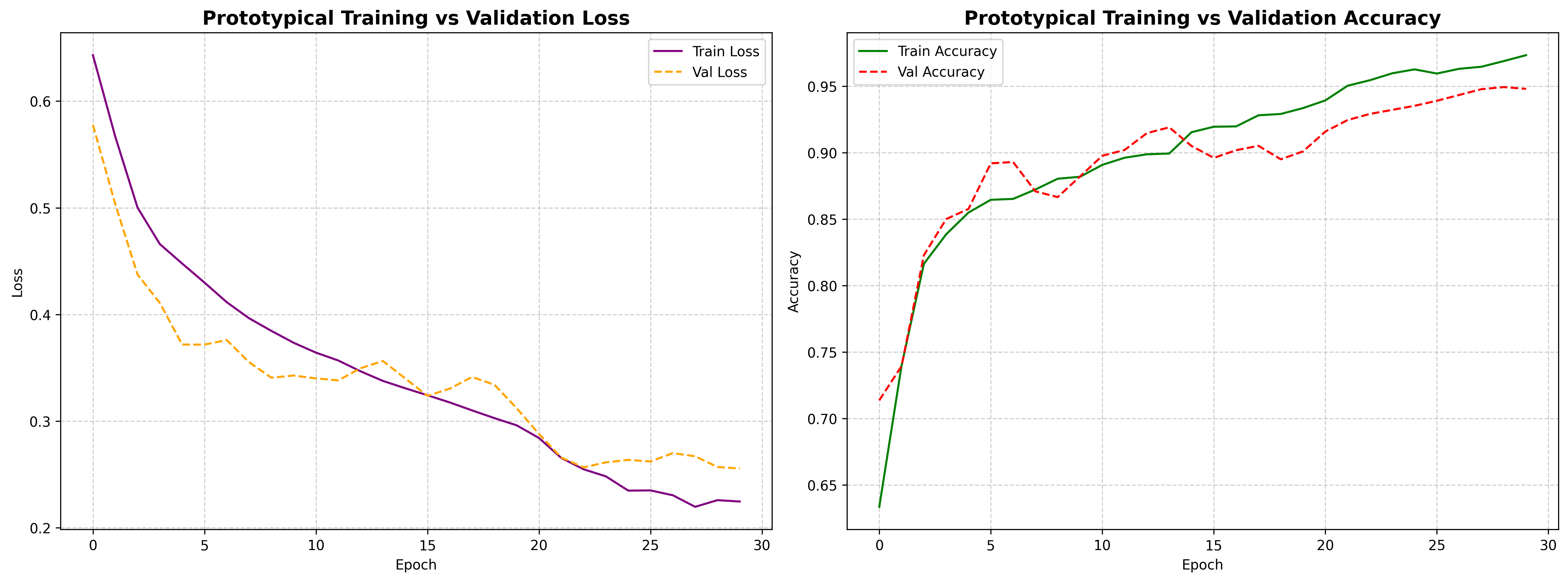}
    \caption{Train-validation loss and accuracy curves for the HLS dataset}
    \label{fig:loss_accuracy_curves}
\end{figure}

Finally, for the \textbf{HLS dataset}, the model achieved 95\% accuracy for the validation set and greater than 97\% for the training set. On the test set, it achieved 95.6\% (as reported in the manuscript-Table 1). The model has shown the highest performance on this dataset, proving its excellent performance capability in noise-free clinical settings.

\subsection{Confusion Matrix}

We have also provided the confusion matrices for the test set of all datasets to demonstrate the true positive rate, true negative rate, false positive rate, and false negative rate.

\newpage

\begin{figure}[ht!]
    \centering
    \includegraphics[width=0.6\textwidth]{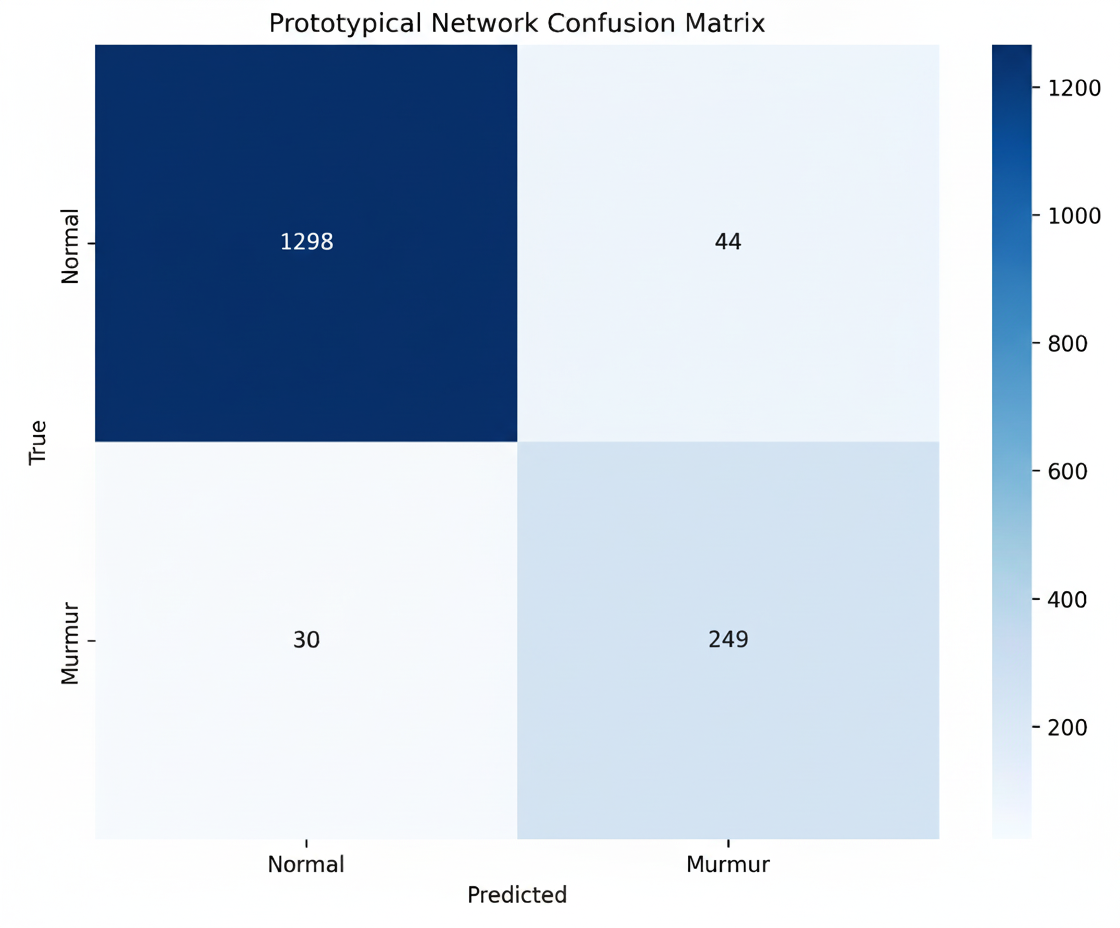}
    \caption{Confusion matrix for the CirCor 2022 dataset}
    \label{fig:confusion_matrix_CirCor}
\end{figure}

For the \textbf{CirCor 2022 dataset}, we maintained the imbalance in the validation and test sets to reflect its actual proportion. Still, the murmur samples were classified correctly with significantly fewer misclassified murmur classes, resulting in a lower recall value, which means the model is effective at identifying murmur-present classes.

\begin{figure}[ht!]
    \centering
    \includegraphics[width=0.6\textwidth]{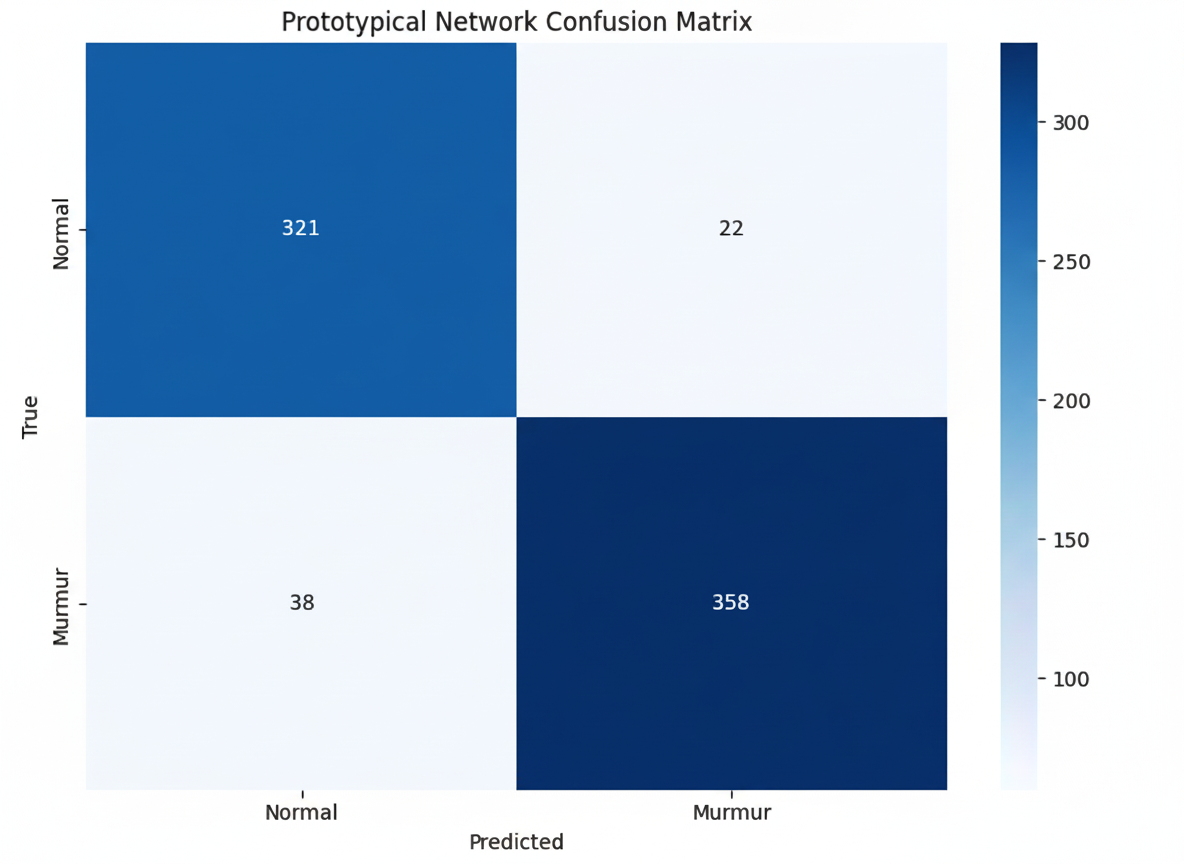}
    \caption{Confusion matrix for the CinC dataset}
    \label{fig:confusion_matrix_CinC}
\end{figure}

For the \textbf{CinC 2016 dataset}, we maintained the imbalance in the validation and test sets according to its actual proportions, where there were more samples of the murmur-present class than the murmur-absent class. As our prototypical network works well with an imbalanced dataset, it emphasizes the minority class over the majority. Therefore, for this dataset, the model prioritized the normal class, which has fewer samples, resulting in fewer misclassifications of the normal class compared to the abnormal class. As a result, the precision value is lower than the recall value. Still, the murmur samples were classified correctly with a significantly lower recall value, indicating that the model is effective at identifying classes with murmurs.

\begin{figure}[ht!]
    \centering
    \includegraphics[width=0.6\textwidth]{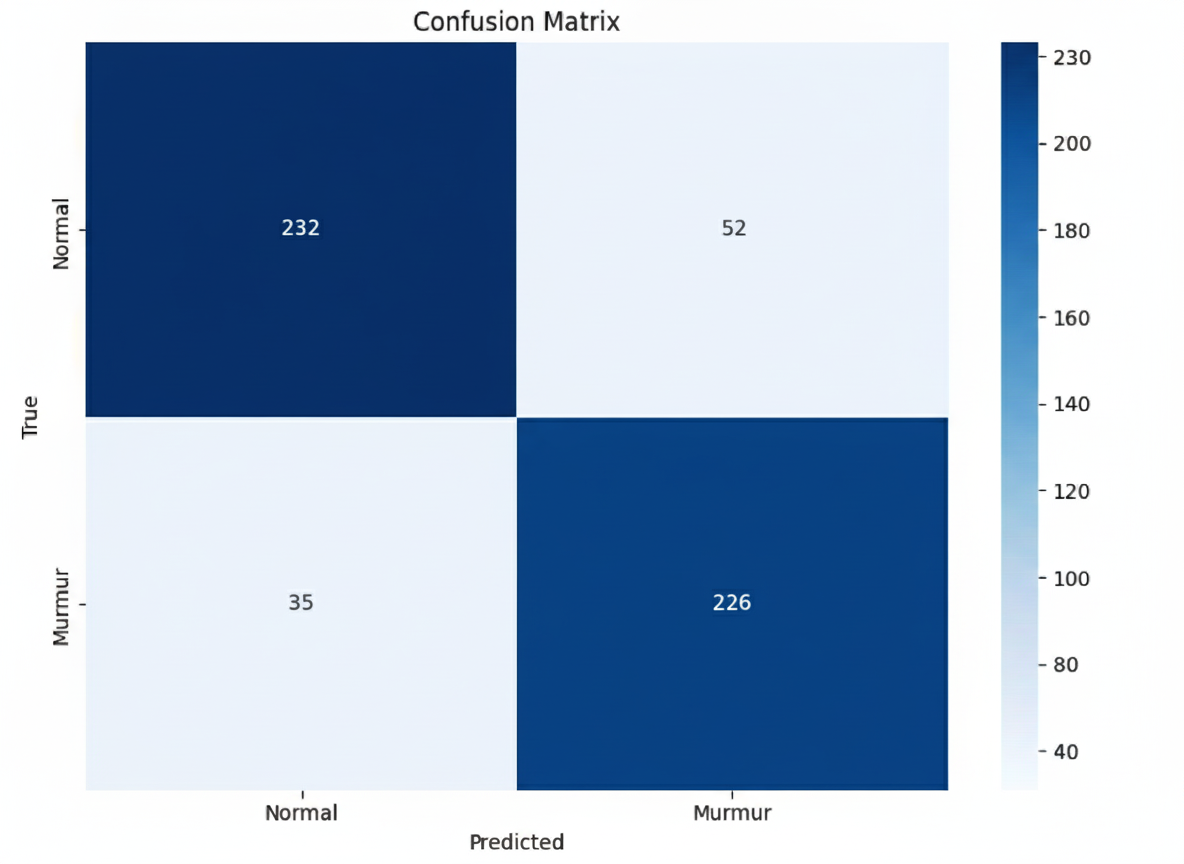}
    \caption{Confusion matrix for the Pascal dataset}
    \label{fig:confusion_matrix_Pascal}
\end{figure}

\newpage

Although the overall performance on the \textbf{Pascal dataset} is lower than that of others, the murmur classes were classified correctly almost 90\% of the time. It generates a higher false negative rate (18\% of the total negative classes) than the previous two datasets, despite being larger, which we consider a setback for our model.

\begin{figure}[ht!]
    \centering
    \includegraphics[width=0.6\textwidth]{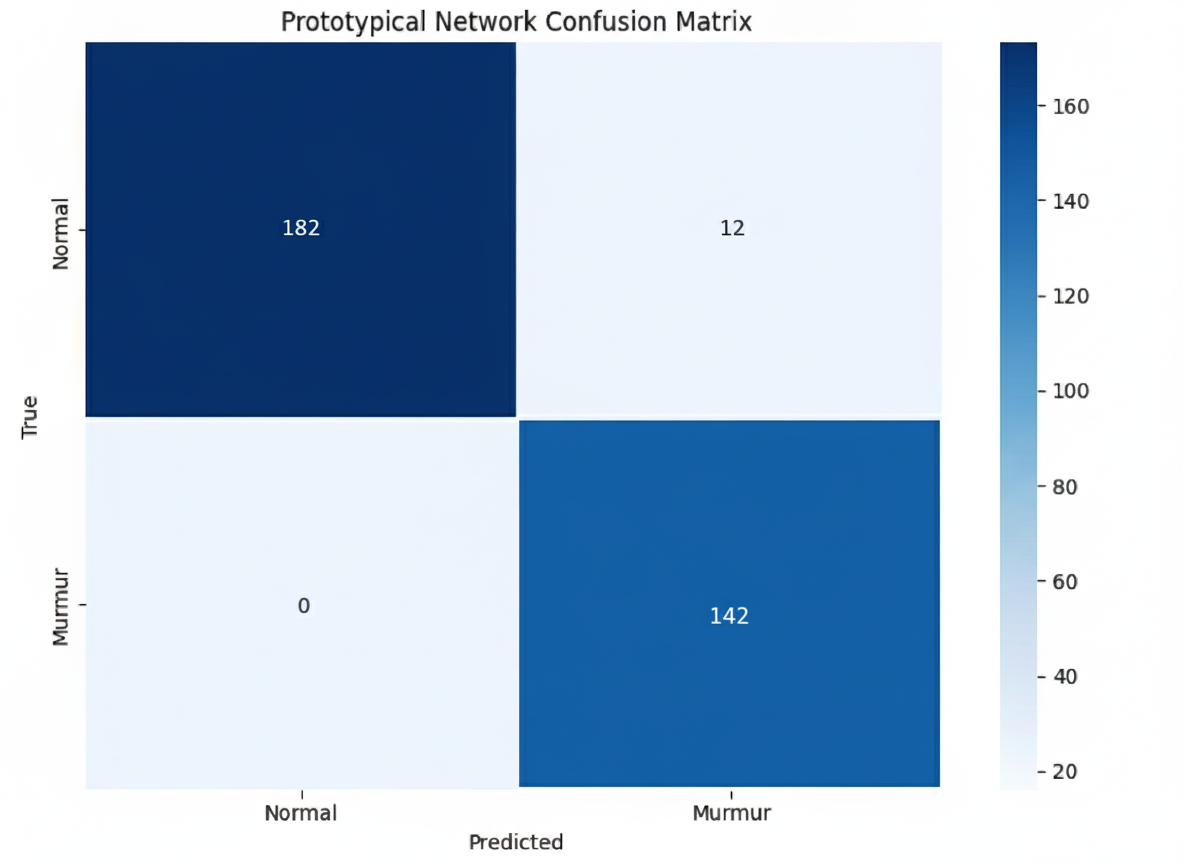}
    \caption{Confusion matrix for the HLS dataset}
    \label{fig:confusion_matrix_HLS}
\end{figure}

Lastly, for the HLS-CMDS dataset, the artificially generated heart murmur sounds showed excellent performance. All the murmur classes were predicted correctly, while a few of the normal classes were also detected as murmurs, which led to a lower precision score than the recall score. Therefore, evaluating this dataset, which contains noise-free, raw, and clear heart murmur recordings, proves our model's better capability of distinguishing heart sounds in a noise-free clinical environment.

\subsection{AUROC and AUPRC Curve}
Finally, we have added the AUROC and AUPRC curves for each dataset. The AUROC curve plots the true positive rate (TPR) against the false positive rate (FPR) across thresholds, measuring overall class separability. An AUPRC curve plots precision against recall across thresholds, focusing on the quality of positive predictions. In our experiment, we emphasized AUPRC because it is more informative than AUROC on imbalanced datasets, as it penalizes false positives and directly reflects performance on the minority class, indicating the reliability of positive predictions.

\begin{figure}[ht!]
    \centering
    \includegraphics[width=\textwidth]{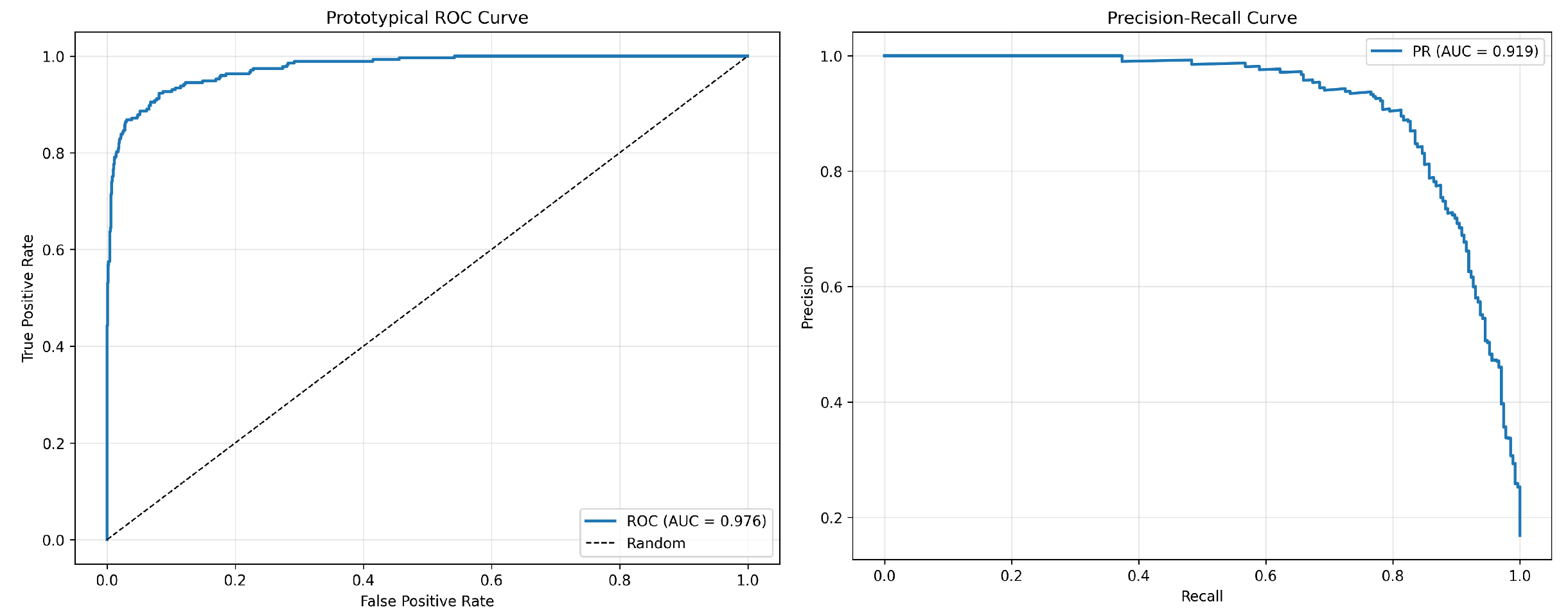}
    \caption{AUROC and AUPRC curves for the CirCor 2022 dataset}
    \label{fig:rocpr_CirCor}
\end{figure}

For the \textbf{CirCor 2022 dataset}, both the ROC and AUPRC tests demonstrated excellent performance. Although there was a slight trade off between precision and recall scores, it has an excellent capability to distinguish the extreme minority murmur class. It achieved high scores for both the Area Under the Receiver Operating Characteristic Curve (AUROC,~97\%) and the Area Under the Precision-Recall Curve (AUPRC,~92\%). Together, these results validate our model as a robust and highly accurate classifier for this task.

\begin{figure}[ht!]
    \centering
    \includegraphics[width=\textwidth]{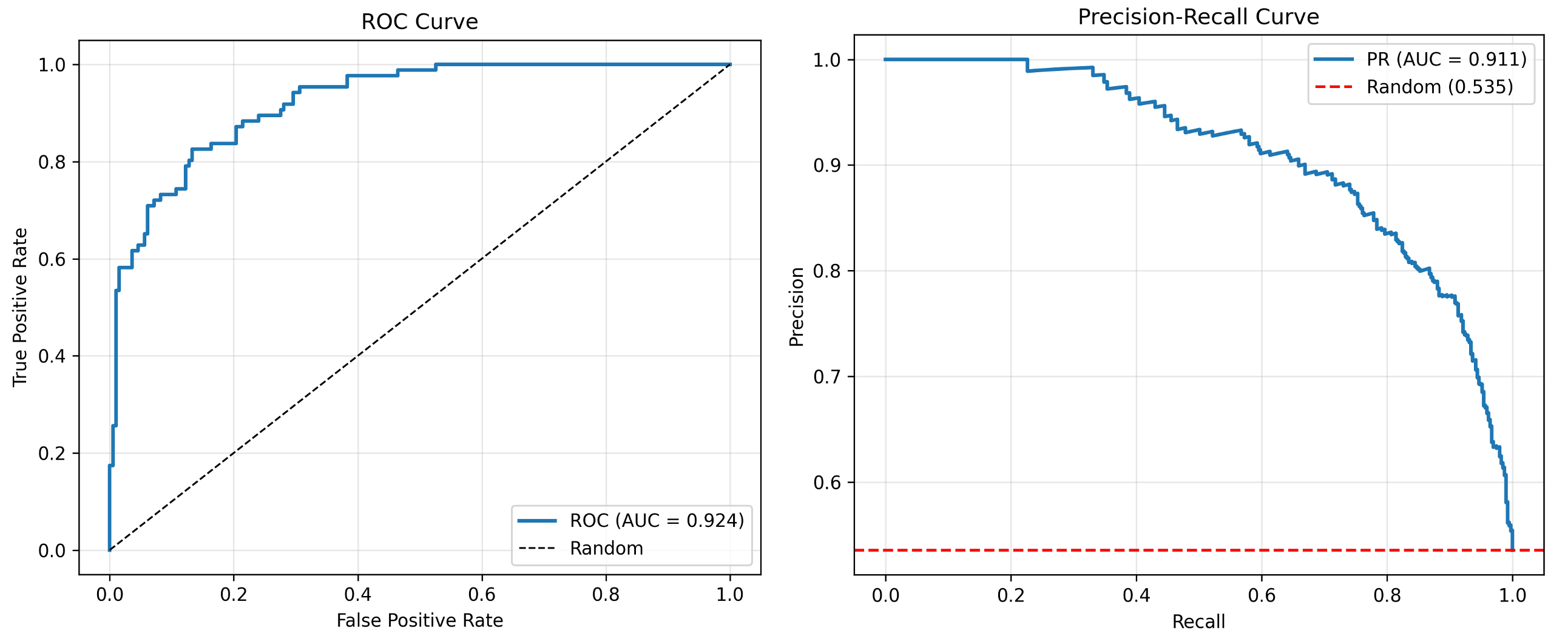}
    \caption{AUROC and AUPRC curves for the CinC 2016 dataset}
    \label{fig:rocpr_CinC}
\end{figure}

The model also demonstrates excellent performance on the \textbf{CinC 2016 dataset}, with an AUROC of 0.924, showing a strong ability to distinguish between positive and negative classes. The AUPRC of 0.911, which is well above the random baseline of 0.535, indicates that the model achieves both high precision and recall in detecting the positive class. Although the PR score is slightly lower than the ROC score (which is typical), it is more meaningful in this case since the positive class prevalence is~53\%. Overall, the model is highly effective and reliable in both balanced and imbalanced evaluation perspectives.

\begin{figure}[ht!]
    \centering
    \includegraphics[width=\textwidth]{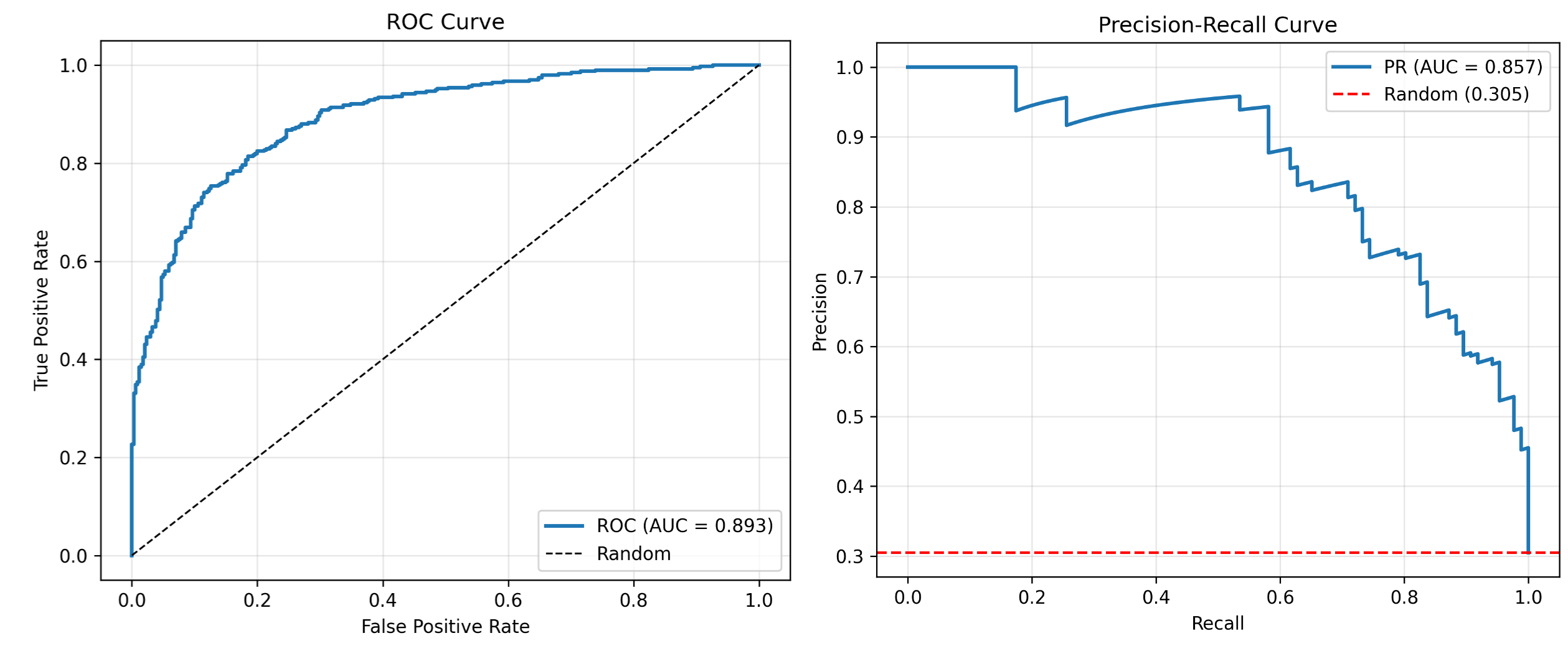}
    \caption{AUROC and AUPRC curves for the Pascal dataset}
    \label{fig:rocpr_Pascal}
\end{figure}

As we have seen, the \textbf{Pascal dataset} yielded comparatively lower scores; the AUROC and AUPRC scores are also lower than those of other datasets. However, the model still exhibits strong performance, with an AUROC of 0.893, indicating that it can effectively separate the positive and negative classes. We intentionally imbalanced the positive and negative classes, and the AUPRC of 0.857 further confirms that it maintains high precision in retrieving positives, significantly above the random baseline of 0.305. Even though the PR score is slightly lower than the ROC score, it is more informative in this imbalanced setting, demonstrating that the model makes reliable positive predictions while effectively handling class imbalance.

\begin{figure}[ht!]
    \centering
    \includegraphics[width=\textwidth]{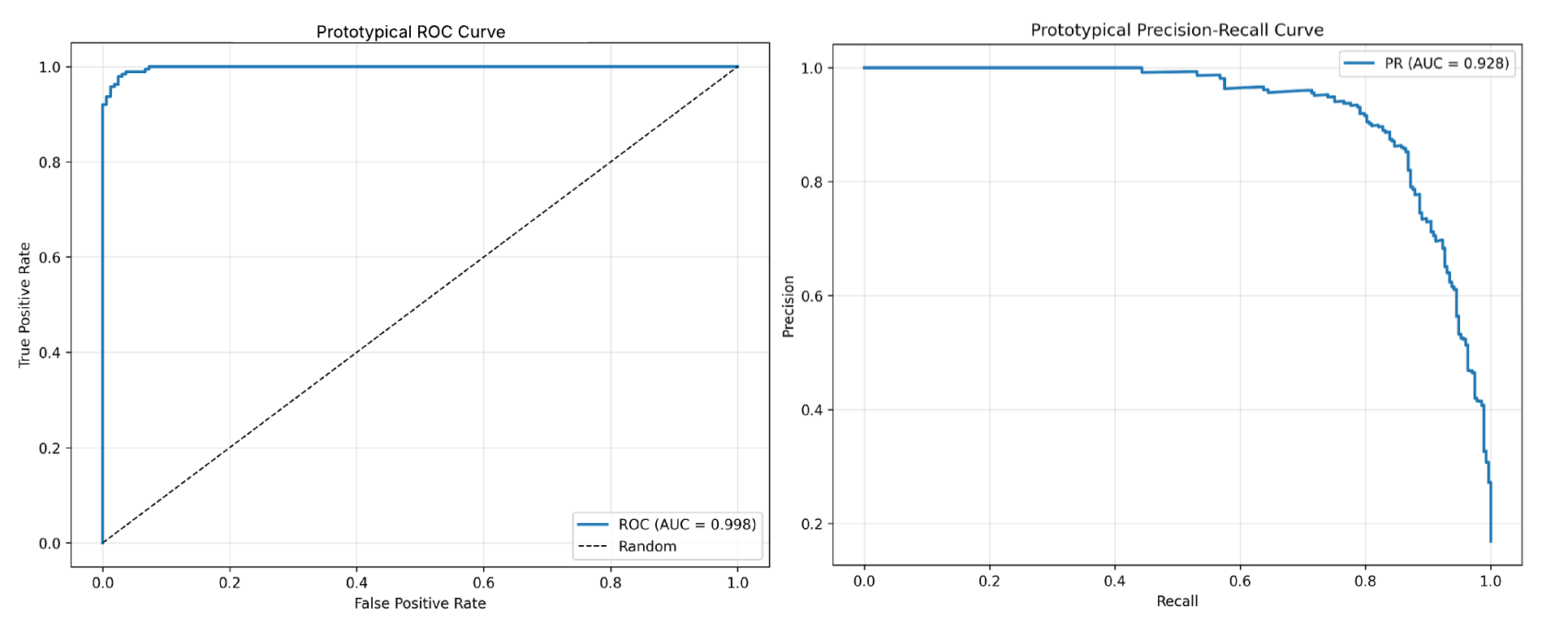}
    \caption{AUROC and AUPRC curves for the HLS-CMDS dataset}
    \label{fig:rocpr_HLS}
\end{figure}

The model demonstrates near-perfect performance on the \textbf{HLS-CMDS dataset}, as it did for the other metrics. This is because the dataset is clean and noise-free, featuring clear synthetic murmur recordings, which demonstrates that the model performs optimally when given a noise-free recording. The AUROC score of 0.998 demonstrates its flawless separation between classes, and an AUPRC of 0.928 confirms that it achieves both high precision and recall. While the PR score is slightly lower than the ROC score, it still indicates that the model makes highly reliable positive predictions, making it very effective for practical use.

% \end{document}

\section*{Funding}
This work is supported by the Research Seed Grant Initiative-2024, BRAC University, Dhaka-1212, Bangladesh.

\section*{Acknowledgements}
The authors express their sincere gratitude to the Biomedical Sciences and Engineering Research Center (BIOSE) at BRAC University for providing the necessary resources, guidance, and support throughout this research.

\section*{Author Contributions}
U.M.M. contributed to the methodology, data curation, formal analysis, prepared the original draft, and developed the visualizations. M.M.H.S. led the conceptualization, conducted the investigation and formal analysis, contributed to the original draft, and managed the project. M.J. handled validation, formal analysis, and writing - review and editing. S.A. provided resources, contributed visualizations, and assisted with writing - review and editing. M.R.H. oversaw supervision and writing - review and editing. M.G.R.A. provided resources, administered the project, secured funding, and supervised the work. All authors reviewed and revised the manuscript and approved the final version for submission.

\section*{Competing Interests}
All authors declare no financial or non-financial competing interests.

\bibliography{sn-bibliography}% common bib file

\end{document}